\documentclass[aps,pra,a4paper,superscriptaddress,longbibliography,reprint]{revtex4-1}

\usepackage[T1]{fontenc}
\usepackage{newtxmath,newtxtext}

\usepackage{amsmath}
\usepackage{amsfonts}
\usepackage{bm}
\usepackage{bbm}
\usepackage{subfigure}
\usepackage{siunitx}

\usepackage{graphicx}
\usepackage{color}
\usepackage{tikz}
\usetikzlibrary{shapes.geometric, arrows}
\usetikzlibrary{calc}
\usetikzlibrary{arrows.meta}
\usetikzlibrary{positioning}
\usetikzlibrary{fit}
\usetikzlibrary{shapes}

\usepackage[acronym,nonumberlist]{glossaries}
\newacronym{DFT}{DFT}{density functional theory}
\newacronym{GAP}{GAP}{Gaussian approximation potential}
\newacronym{SOAP}{SOAP}{smooth overlap of atomic positions}
\newacronym{MD}{MD}{molecular dynamics}
\newacronym{RMSE}{RMSE}{root mean squared error}
\newacronym{MAE}{MAE}{mean absolute error}
\newacronym{ML}{ML}{machine learning}
\newacronym{MLP}{MLP}{\gls{ML} potential}
\newacronym{PBE}{PBE}{Perdew-Burke-Ernzerhof}
\newacronym{KS}{KS}{Kohn-Sham}
\newacronym{vdW}{vdW}{van der Waals}
\newacronym{TS}{TS}{Tkatchenko-Scheffler}
\newacronym{MBD}{MBD}{many-body dispersion}
\newacronym{ATM}{ATM}{Axilrod-Teller-Muto}
\newacronym{CFDM}{CFDM}{the coupled fluctuating dipole model}
\newacronym{SCS}{SCS}{self-consistently screened}
\newacronym{ACFDT}{ACFDT}{the adiabatic connection fluctuation-dissipation theorem}
\newacronym{RPA}{RPA}{the random-phase approximation}
\newacronym{XDM}{XDM}{exchange-hole dipole moment}

\begin{document}

\title{Atom-wise formulation of the many-body dispersion problem for
linear-scaling van der Waals corrections}

\author{Heikki Muhli}
\email{heikki.muhli@gmail.com}
\affiliation{Department of Applied Physics,
Aalto University, Konemiehentie 1, 02150, Espoo, Finland}
\affiliation{Department of Chemistry and Materials Science,
Aalto University, Kemistintie 1, 02150 Espoo, Finland}

\author{Tapio Ala-Nissila}
\affiliation{Department of Applied Physics, QTF Center of Excellence, Aalto University, P.O. Box 15600, FIN-00076 Aalto, 
Espoo, Finland}
\affiliation{Interdisciplinary Centre for Mathematical Modelling
and Department of Mathematical Sciences, Loughborough University,
Loughborough, Leicestershire LE11 3TU, United Kingdom}

\author{Miguel A. Caro}
\email{mcaroba@gmail.com}
\affiliation{Department of Chemistry and Materials Science,
Aalto University, Kemistintie 1, 02150 Espoo, Finland}

\date{9 July 2024}

\begin{abstract}
A common approach to modeling dispersion interactions and overcoming the inaccurate description of long-range correlation effects in electronic structure calculations is the use of pairwise-additive potentials, as in the Tkatchenko-Scheffler [Phys. Rev. Lett. \textbf{102}, 073005 (2009)] method. In previous work [Phys. Rev. B \textbf{104}, 054106 (2021)], we have shown how these are amenable to highly efficient atomistic simulation by machine learning their local parametrization. However, the atomic polarizability and the electron correlation energy have a complex and non-local many-body character and some of the dispersion effects in complex systems are not sufficiently described by these types of pairwise-additive potentials. Currently, one of the most widely used rigorous descriptions of the many-body effects is based on the many-body dispersion (MBD) model [Phys. Rev. Lett. \textbf{108}, 236402 (2012)]. In this work, we show that the MBD model can also be locally parametrized to derive a local approximation for the highly non-local many-body effects. With this local parametrization, we develop an atom-wise formulation of MBD that we refer to as linear MBD (lMBD), as this decomposition enables linear scaling with system size. This model provides a transparent and controllable approximation to the full MBD model with tunable convergence parameters for a fraction of the computational cost observed in electronic structure calculations with popular density-functional theory codes. We show that our model scales linearly with the number of atoms in the system and is easily parallelizable.
Furthermore, we show how using the same machinery already established in previous work for predicting Hirshfeld volumes with machine learning enables access to large-scale simulations with MBD-level corrections.
\end{abstract}

\maketitle

\section{Introduction}

In quantum-mechanical electronic-structure calculations, the practical approach to solve the many-body problem of $N$ electrons is often reformulated as a problem of $N$ effective independent particles. In \gls{DFT}, the \gls{KS}~\cite{kohn1965self} reference system is one such example. This kind of approaches are often capable of capturing more than 99~\% of the total electronic energy~\cite{stohr2019theory}. However, the remaining part of the total energy can play an important role for many important observables of the system, such as relative energies and binding properties. An example of the importance of this remaining part is an Argon dimer: a \gls{KS} calculation with the PBE0~\cite{perdew1996rationale,adamo1999toward} hybrid functional captures 99.95~\% of the total energy but only 15~\% of the interaction energy of the dimer~\cite{stohr2019theory}. The missing part is contributed by the correlated motion of the electrons. The main component of this long-range correlation energy is known as the \gls{vdW} dispersion interaction.

There are various ways to incorporate dispersion interactions into a \gls{DFT} calculation with varying levels of accuracy and computational cost. These vary from explicit non-local density functionals to pairwise and many-body corrections that are added to the \gls{DFT} calculations in post-processing. The non-local density functionals, such as vdW-DF and optB88, rely on approximate formulations with added physical constraints to include long-range correlations~\cite{PhysRevLett.92.246401,sato2005van,PhysRevLett.76.102,PhysRevB.82.081101,klimevs2009chemical}. However, these approaches are not useful for large systems because of their high computational cost~\cite{otero2020many}. 

A separate long-range correction that is added on top of the baseline \gls{DFT} energy, where the functional itself does not capture long-range correlations, allows for a much larger system and is also better suited for the purpose of machine learning these interactions. 
The pairwise-additive dispersion correction is the best known one. It was originally formulated by London in 1930~\cite{london1930theorie} and, indeed, many of the currently used pairwise-additive dispersion formulas still follow the familiar $1/r^6$ relationship, i.e., the pairwise dispersion energies depend inversely on the sixth power of the interatomic distance:
\begin{equation}
    E_{\text{disp}} = -\frac{1}{2}\sum\limits_{i}\sum\limits_{j\neq i} \frac{C_{6,ij}}{r_{ij}^6},
    \label{london-dispersion}
\end{equation}
where the sums go over all atoms and $C_{6,ij}$ is the dipole-dipole dispersion coefficient that can be defined in varying ways. Some examples of these corrections include Grimme's D2 method~\cite{grimme2006semiempirical}, where the coefficients are derived from free atomic values, Grimme's D3 method~\cite{grimme2010consistent}, where an $1/r^8$ term is added and the coefficients also depend on the atomic coordination numbers, and the \gls{TS} approach~\cite{tkatchenko2009accurate}, where the polarizabilities are first scaled with the partitioned electronic density and then integrated to obtain the coefficients, following the Casimir-Polder formulation~\cite{PhysRev.73.360}. This method to obtain the coefficients was later also adopted by Grimme's D4 method~\cite{caldeweyher2019generally}. 

Grimme {\it et al.} also proposed higher-than-two-body terms for dispersion corrections in the context of \gls{DFT}~\cite{grimme2010consistent}. The added term, known as \gls{ATM} term~\cite{axilrod1943interaction,muto1943force}, is an additional three-body correction to account for effects that are not captured by a pairwise-additive model, and is reported to typically contribute less than $5$--$10$~\% of the dispersion energy~\cite{grimme2010consistent}. However, due to the triple sum instead of double sum, as seen in Eq.~\eqref{london-dispersion}, the computational cost for calculating this term increases from $O(N^2)$ to $O(N^3)$ scaling with the number of atoms $N$ in the system. Furthermore, the three-body term could even produce the wrong sign for the dispersion energy in some cases~\cite{tkatchenko2012accurate}. This significant increase in the computational cost and the questionable increase in the accuracy means that the method has not been as widely adopted as the pairwise approximations.

To improve the accuracy of the dispersion correction, Tkatchenko {\it et al.} derived a method from \gls{CFDM}~\cite{cole2009nanoscale} to calculate accurate dispersion corrections to infinite body order~\cite{tkatchenko2012accurate}. This new method, called \gls{MBD}, also used polarizabilities that are \gls{SCS}~\cite{oxtoby1975collisional,thole1981molecular,felderhof1974propagation} with respect to other atoms in the system, and can be used to improve the pairwise \gls{TS} correction~\cite{tkatchenko2012accurate}. \gls{MBD} reaches remarkably good accuracy when compared to coupled-cluster results~\cite{tkatchenko2012accurate}, which are often considered of benchmark quality. \gls{MBD} incurs a significant computational cost for matrix diagonalization, where the dimensions of the matrix are given by the size of the system (number of atoms). The method was later rederived, also by Tkatchenko {\it et al.}, from \gls{ACFDT} and \gls{RPA}~\cite{dobson2012calculation,tkatchenko2013interatomic}. \gls{ACFDT}-\gls{RPA} can also be used to directly calculate electronic correlation energies (which include dispersion) but the computational cost is extreme because of the need to evaluate the \gls{KS} response function matrix, among other things, and thus not suitable for large systems~\cite{dobson2012calculation,yeh2023low,PhysRevB.109.035103,ambrosetti2014long}. The \gls{ACFDT}-\gls{RPA} method also suffers from over-correlation at short range~\cite{dobson2012calculation}, which can be solved by range-separating the interaction in the derivation of the \gls{MBD} model~\cite{tkatchenko2013interatomic,buvcko2016many,ambrosetti2014long}. Furthermore, local quantities, that could be taken advantage of with \gls{ML} thus potentially avoiding an expensive electronic structure calculation at every step, cannot be used within the \gls{RPA} approach. On the other hand, \gls{MBD} allows this by casting the problem into a system of quantum harmonic oscillators~\cite{bereau_2018}.

A different perspective by Otero de la Roza {\it et al.}~\cite{otero2020many} offers criticism towards the inclusion of unnecessary \textit{atomic} many-body effects when the same or better accuracy could be achieved by including more pairwise-additive dispersion terms where the coefficients have been calculated using the \gls{XDM} method~\cite{becke2007exchange}, taking into account the more important \textit{electronic} many-body effects. This method relies on the concept of electrons leaving behind exchange holes when they travel through the system, inducing dipoles and higher order multipoles~\cite{becke2007exchange,otero2020many}. These exchange holes can be calculated semi-locally~\cite{otero2020many,PhysRevA.39.3761} and then used to calculate the dispersion coefficients using moment integrals~\cite{otero2020many,becke2007exchange}.

The proponents of \gls{XDM} argue that many of the dispersion models use arbitrarily defined damping functions and empirical parameters and then correct the part of the dispersion missing because of these definitions with higher-order atomic many-body terms. According to Otero de la Roza {\it et al.}, these computationally expensive higher-order terms would not be necessary if the electronic many-body (i.e., the construction of the dispersion coefficients) were carried out more rigorously. Their claim, backed by calculations~\cite{otero2020many,price2023xdm}, is that electronic many-body effects can result in a 50~\% variation for the leading coefficients in pairwise-additive dispersion models, while the atomic many-body terms represent less than 1~\% of the dispersion energy, making their contribution almost negligible for anything but noble-gas trimers. Their analysis seems to only involve the \gls{ATM} term for atomic many-body effects which is usually the leading term beyond pairwise-additive dispersion. However, as we already mentioned, the inclusion of only the \gls{ATM} term can result in the wrong sign for the dispersion energy~\cite{tkatchenko2012accurate} and this term might not be the leading beyond-pairwise term in some systems, where the odd terms are zero by symmetry. Nevertheless, the \gls{XDM} method offers an interesting alternative for calculating dispersion corrections with many-body effects. The method itself is attractive because it only requires computationally cheap pairwise terms for the atomic many-body effects, and the semi-local calculation of the exchange holes might allow a \gls{ML} approach, given that the method is ``local enough''. In our approach, we have decided to focus on the \gls{MBD} model of Tkatchenko {\it et al.}~\cite{tkatchenko2012accurate,tkatchenko2013interatomic,buvcko2016many} instead. \gls{MBD} effectively includes both the electronic and atomic many-body effects by construction but has a higher computational cost than the \gls{XDM} model.

The \gls{MBD} formalism derived from \gls{ACFDT}-\gls{RPA} is also used by the \gls{DFT} code VASP~\cite{buvcko2016many,kresse1993ab,kresse1996efficiency,kresse1996efficient} and we will describe it in more detail in the following section. We have chosen to use this formalism because it provides range separation out of the box~\cite{ambrosetti2014long,buvcko2016many}, relies on local quantities we have already successfully used in \gls{ML} context before~\cite{muhli2021machine}, and expresses the energy with an explicit dependence on the atomic positions and polarizabilities. 
In addition, VASP is by far the most widely used electronic-structure code for periodic calculations, and so it provides a relevant benchmark \gls{MBD} implementation for accuracy and computational cost comparisons.

There exist other recent implementations of \gls{MBD}; however, they differ significantly from the present work. Poier {\it et al.} published a code that is capable of calculating \gls{MBD} energies for large systems and the code scales linearly with the number of atoms in the system~\cite{poier2022n}. This implementation calculates the \gls{MBD} energies stochastically, using a Lanczos trace estimator~\cite{ubaru2017fast}. The idea itself is interesting, but the reported standard deviations of the stochastic approach seem large for dispersion energies. Furthermore, the stochastic approach does not allow for the analytic calculation of forces, which are important for interatomic force fields. Another recent development is the libMBD library by Hermann {\it et al.}, written in Fortran with bindings to C and Python~\cite{hermann2023libmbd}. It offers functionality to calculate \gls{MBD} energies and forces in an efficient and flexible form. The code makes no assumptions about the method that was used to obtain the input polarizabilities or Hirshfeld volumes~\cite{hirshfeld1977bonded} required for the calculation, and thus readily also offers support for \gls{ML}-derived parametrizations. However, the asymptotic scaling of the method is still $O(N^3)$ (or $O(N^2)$ for smaller systems where the matrix construction dominates)~\cite{hermann2023libmbd}. While both of these recent implementations are certainly useful, they do not offer what we need for large-scale \gls{MD} simulations: linear-scaling and parallelizable \gls{MBD} correction that is also capable of calculating the \gls{MBD} forces within a controllable approximation. To this end, in the present work we show how this can be done with an atom-wise formulation.

\section{Background}

In this section we will first discuss the key concept of ``non-additivity'' in the context of \gls{MBD}. Non-additivity is the main reason there is no atom-wise formulation of the \gls{MBD} method to date. We will describe the three types of non-additivity that can arise in dispersion models, which of them are present in \gls{MBD}, and which of them we can address in this work. Then we will give a brief but detailed description of the implementation of \gls{MBD} that is used by VASP~\cite{kresse1993ab,kresse1996efficiency,kresse1996efficient,buvcko2016many}, and was originally derived from \gls{ACFDT}-\gls{RPA}~\cite{dobson2012calculation} by Tkatchenko {\it et al.}~\cite{tkatchenko2013interatomic}.

\subsection{Types of non-additivity}

Dobson~\cite{dobson2014beyond} defines three different types of non-additivity related to dispersion: type A, type B and type C. Type A simply states that it would not be sensible to use an interaction derived for free atoms to describe a system where the atoms are bonded. Almost all modern theories take this into account, including most pairwise-additive models.

Type-B non-additivity is the most important one in this context. It follows from classical electromagnetism that an additional polarizable center can screen the interaction between two atoms, changing the correlation energy~\cite{dobson2014beyond}. Here Dobson mentions the \gls{MBD} method and how the dispersion energies calculated with it for $N$ atomic centers can be completely different from the ones calculated with pairwise-additive methods, where the dispersion energy can be seen as proportional to $N^2$~\cite{dobson2014beyond}. Because of type-B non-additivity, the \gls{MBD} energy cannot be written as a sum of pairwise-additive terms. However, we will show later that the energy can be written in terms of \emph{atomic} contributions and we will refer to this model as ``additive'' in the text, because it allows us to write the dispersion energy as a sum of local energies. Approximating this type-B non-additivity with an additive model is one of the main results in this work.

Finally, type-C non-additivity is attributed to the electronic delocalization~\cite{dobson2014beyond}. In atom-based models, each electron is assumed to ``belong'' to a certain atom in some sense. This is also the case with \gls{MBD}, where the underlying effective Hirshfeld volumes give the electronic density of the atom in a molecule compared to the free atom. Models that rely on \textit{local} parametrization of atom-wise properties, such as effective Hirshfeld volumes, are fundamentally unable to overcome type-C non-additivity: type C can be only tackled properly with \gls{RPA}~\cite{dobson2012calculation}. Fortunately, a large portion of the type-C effects are suppressed by type-B screening for most systems~\cite{dobson2014beyond}.

\subsection{Many-body dispersion in density-functional theory}

In the following, we describe the calculation of MBD energies using the non-local formalism of VASP~\cite{kresse1993ab,kresse1996efficiency,kresse1996efficient}. Later, in Sec.~\ref{localization}, we describe the local approximation we have derived from the full non-local model. In \gls{DFT} calculations, the \gls{MBD} energy of a system containing $N$ atoms is given by (using Hartree units throughout the text for simplicity)~\cite{buvcko2016many}:
\begin{equation}
    E_{\text{MBD}} = \int\limits_0^\infty \frac{\text{d}\omega}{2\pi} \text{Tr}\{ \text{ln}(\mathbf{1}-\mathbf{A}_{\text{LR}}(\omega)\mathbf{T}_{\text{LR}})\},
    \label{E_MBD}
\end{equation}
where the integration is over frequencies, $\text{ln}(\mathbf{X})$ is a matrix logarithm of matrix $\mathbf{X}$, $\mathbf{A}_{\text{LR}} \in \mathbb{R}^{3N \times 3N}$ is the diagonal long-range polarizability matrix containing the isotropic frequency-dependent atomic polarizabilities of each atom on the diagonal (three times each) and $\mathbf{T}_{\text{LR}} \in \mathbb{R}^{3N \times 3N}$ is the frequency-independent long-range dipole interaction tensor. The elements of $\mathbf{A}_{\text{LR}}$ are given by:
\begin{equation}
    A_{\text{LR},ij}^{\alpha\beta} (\omega) = \delta_{ij}\delta_{\alpha\beta}\tilde{\alpha}^{\text{iso}}_i(\omega),
\end{equation}
where $\delta_{ij}$ is a Kronecker delta, $i$ and $j$ denote the atomic indices, and $\alpha$ and $\beta$ denote the Cartesian components of the atomic contributions. $\tilde{\alpha}^{\text{iso}}_i(\omega)$ is the isotropic frequency-dependent polarizability that is self-consistently solved for the whole system taking into account the initial polarizabilities and the dipole couplings. 
The elements of $\mathbf{T}_{\text{LR}}$ are given by:
\begin{equation}
    T_{\text{LR},ij}^{\alpha\beta} = f(r_{ij},\tilde{S}_{\text{vdW},ij}) T_{ij}^{\alpha\beta},
    \label{T_LR}
\end{equation}
where $f(r_{ij},\tilde{S}_{\text{vdW},ij})$ is the damping function that is used to screen the dipole coupling tensor $T_{ij}^{\alpha\beta}$ to obtain the long-range contribution. The Fermi-type damping function is given by:
\begin{equation}
   f(r_{ij},\tilde{S}_{\text{vdW},ij}) = \frac{1}{1+\text{exp}[-d(r_{ij}/\tilde{S}_{\text{vdW},ij} - 1)]},
   \label{damping-function}
\end{equation}
where $d = 6$ is a fixed parameter~\cite{buvcko2016many}, $r_{ij}$ is the interatomic distance between atoms $i$ and $j$, and $\tilde{S}_{\text{vdW},ij}$ is the scaled \gls{vdW} radius between atoms $i$ and $j$:
\begin{equation}
    \tilde{S}_{\text{vdW},ij} = \beta(\tilde{R}_{\text{vdW},i} + \tilde{R}_{\text{vdW},j}).
    \label{Svdw}
\end{equation}
Here the parameter $\beta$ is optimized for the used density functional; for PBE the value is $\beta = 0.83$~\cite{buvcko2016many}. Each variable here denoted with a tilde, such as $\tilde{S}_{\text{vdW},ij}$, is obtained after the self-consistency cycle and depends on the static isotropic polarizabilities $\tilde{\alpha}^{\text{iso}}_i(0)$. The elements of the full-rank second-order dipole interaction tensor are given by:
\begin{equation}
    T_{ij}^{\alpha\beta} = \frac{\partial}{\partial r_{ij}^{\alpha}}\frac{\partial}{\partial r_{ij}^{\beta}} \Bigg(\frac{1}{r_{ij}}\Bigg) = \frac{3r_{ij}^{\alpha}r_{ij}^{\beta}-\delta_{\alpha\beta}r_{ij}^2}{r_{ij}^5},
\end{equation}
where $r_{ij}^{\alpha}$ are the Cartesian components of the interatomic position vector $\mathbf{r}_{ij}$.

The self-consistent polarizabilities for the short-range screened atom in a molecule are given by the SCS equation~\cite{tkatchenko2012accurate,buvcko2016many}:
\begin{equation}
    \boldsymbol{\tilde{\alpha}}_i (\omega) = \bar{\alpha}_i (\omega)\Big(\mathbf{1}-\sum\limits_{j\neq i}^N \mathbf{T}_{\text{SR},ij}(\omega) \boldsymbol{\tilde{\alpha}}_j(\omega)\Big) \in \mathbb{R}^{3\times 3}.
    \label{SCS}
\end{equation}
The isotropic polarizabilites $\tilde{\alpha}^{\text{iso}}_i(\omega)$ are given as one third of the trace of these polarizabilities. In practice, the SCS equation is solved for every atom at once by the matrix equation:
\begin{equation}
    \tilde{\boldsymbol{{\alpha}}} (\omega) = \tilde{\mathbf{B}} (\omega)^{-1} \mathbf{d},
    \label{polarizabilities}
\end{equation}
where $\tilde{\boldsymbol{\alpha}} \in \mathbb{R}^{3N \times 3}$ is the many-body polarizability matrix and $\tilde{\mathbf{B}}$ is given by (see Eq.~\eqref{SCS}):
\begin{equation}
    \tilde{B}_{ij}^{\alpha\beta}(\omega) = \frac{1}{\bar{\alpha}_i(\omega)}\delta_{ij}\delta_{\alpha\beta} + T_{\text{SR},ij}^{\alpha\beta} (\omega).
    \label{B_eq}
\end{equation}
The matrix $\mathbf{d} \in \mathbb{R}^{3N \times 3}$ consists of identity matrix blocks $\mathbf{d}_i = \mathbf{I}_3 \in \mathbb{R}^{3\times 3}$ for each atom. The isotropic polarizability is given by the average of the trace of the corresponding $3 \times 3$ subblock:
\begin{equation}
    \tilde{\alpha}^{\text{iso}}_i(\omega) = \frac{1}{3}\mathrm{Tr}\{\boldsymbol{\tilde{\alpha}}_i (\omega)\} = \frac{1}{3} \sum\limits_{\alpha} \sum\limits_{\beta} \sum\limits_{j=1}^N [\mathbf{\tilde{B}}(\omega)^{-1}]_{ij}^{\alpha\beta} d_j^{\beta\alpha}.
\end{equation}
Next, we will fully define $\tilde{B}_{ij}^{\alpha\beta}$ by defining $\bar{\alpha}_i$ and $T_{\text{SR},ij}^{\alpha\beta}$~\cite{buvcko2016many}:
\begin{equation}
    \bar{\alpha}_i(\omega) = \frac{\alpha_i^{\text{atom}}\nu_i}{1+\Big(\frac{\omega}{\omega_i}\Big)^2},
    \label{bar-alpha}
\end{equation}
with $\omega_i = {4C_{6,ii}^{\text{atom}}}/[{3(\alpha_i^{\text{atom}})^2}]$, $\alpha^{\text{atom}}_i$ and $C_{6,ii}^{\text{atom}}$ being the free atom polarizabilities and dispersion coefficients. Here, $\nu_i$ is the effective Hirshfeld volume~\cite{hirshfeld1977bonded} of atom $i$; it gives the effective volume the electrons occupy in a bonded atom as compared to the corresponding volume of the free atom. 

The elements of the short-range dipole interaction tensor are given by~\cite{buvcko2016many}:
\begin{multline}
    T_{\text{SR},ij}^{\alpha\beta}(\omega) = (1-f(r_{ij},S_{\text{vdW},ij})) \times \\ \frac{\partial}{\partial r_i^{\alpha}} \frac{\partial}{\partial r_j^{\beta}} \Bigg(\frac{\text{erf}(r_{ij}/\sigma_{ij}(\omega))}{r_{ij}}\Bigg) 
    = (1-f(r_{ij},S_{\text{vdW},ij})) \times \\ \Bigg\{ -T_{ij}^{\alpha\beta} g(r_{ij},\sigma_{ij}(\omega)) + h(r_{ij},\sigma_{ij}(\omega))  \Bigg\}.
    \label{T_SR}
\end{multline}
Note that the damping function uses the \gls{vdW} radii without self-consistent screening, that is, we use $S_{\text{vdW},ij}$ instead of $\tilde{S}_{\text{vdW},ij}$ here. The functions $g(r_{ij},\sigma_{ij}(\omega))$ and $h(r_{ij},\sigma_{ij}(\omega))$ are defined as:
\begin{equation}
    g(r_{ij},\sigma_{ij}(\omega)) = \text{erf}(r_{ij}/\sigma_{ij}(\omega)) - \frac{2}{\sqrt{\pi}} \frac{r_{ij}}{\sigma_{ij}(\omega)}\text{exp}\left(-\frac{r_{ij}^2}{\sigma_{ij}^2}\right),
    \label{g-func}
\end{equation}
\begin{equation}
    h(r_{ij},\sigma_{ij}(\omega)) = \frac{4}{\sqrt{\pi}} \Bigg(\frac{r_{ij}}{\sigma_{ij}(\omega)}\Bigg)^3\Bigg(\frac{r_{ij}^\alpha r_{ij}^{\beta}}{r_{ij}^5}\Bigg)\text{exp}\left(-\frac{r_{ij}^2}{\sigma_{ij}^2}\right),
    \label{h-func}
\end{equation}
and the frequency-dependent attenuation length as:
\begin{equation}
    \sigma_{ij}(\omega) = \sqrt{\sigma_i(\omega)^2 + \sigma_j(\omega)^2}
\end{equation}
with
\begin{equation}
    \sigma_i(\omega) = \Bigg( \sqrt{\frac{2}{\pi}} \frac{\bar{\alpha}_i(\omega)}{3} \Bigg)^{1/3}.
\end{equation}
Here we define the initial uncoupled vdW radii using the \gls{TS} method~\cite{tkatchenko2009accurate} and the effective Hirshfeld volumes:
\begin{equation}
    S_{\text{vdW},ij} = \beta(R_{\text{vdW},i} + R_{\text{vdW},j})
\end{equation}
and 
\begin{equation}
    R_{\text{vdW},i} = R_{\text{vdW},i}^{\text{atom}} \nu_i^{1/3},
\end{equation}
where $R_{\text{vdW},i}^{\text{atom}}$ is the free-atom value. Once the SCS polarizabilities have been solved, we can calculate the \gls{SCS} radii with:
\begin{equation}
    \tilde{R}_{\text{vdW},i} = R_{\text{vdW},i}^{\text{atom}} \Bigg(\frac{\tilde{\alpha}_i^{\text{iso}}}{\alpha_i^{\text{atom}}}\Bigg)^{1/3}.
    \label{scaled-rvdw}
\end{equation}

\section{Methodology}
\label{localization}

Next, we describe our local linear-scaling approach to \gls{MBD}, which we refer to as lMBD. This is necessary to keep the computational effort tractable in atomistic simulations with \gls{ML} force fields that are based on local structural descriptors. Although our method is agnostic with respect to how the Hirshfeld volumes were derived (e.g., they could be obtained from a \gls{DFT} calculation), to test it for large systems we rely on efficient \gls{ML}-based prediction~\cite{muhli2021machine}. In our \gls{ML} model, the Hirshfeld effective volume $\nu_i$ is the only quantity that is predicted using kernel regression with many-body \gls{SOAP} descriptors~\cite{bartok2013representing,caro2019optimizing} (similar to the \gls{GAP}~\cite{bartok2010gaussian,bartok2015gaussian} framework):
\begin{equation}
    \nu_i = \sum\limits_{s\in S} \alpha_s (\mathbf{q}_s \mathbf{q}_i^T)^\zeta,
    \label{kernel-expression}
\end{equation}
where $S$ denotes the sparse set of reference atomic structures, $\mathbf{q}_i$ are \gls{SOAP} descriptors and $\alpha_s$ are the fitting coefficients obtained from the \gls{ML} model. $\zeta$ is an empirical parameter that is used to adjust the ``sharpness'' of the kernel. Using these predicted Hirshfeld effective volumes, all other quantities can be obtained analytically from the local approximation.

\subsection{Local energies}

\tikzstyle{startstop} = [rectangle, rounded corners, 
minimum width=3cm, 
minimum height=1cm,
text centered, 
draw=black, 
fill=red!30]

\tikzstyle{io} = [trapezium, 
trapezium stretches=true,
trapezium left angle=70, 
trapezium right angle=110, 
minimum width=2cm, 
minimum height=1cm, text centered,
text width=3cm,
draw=black, fill=blue!30]

\tikzstyle{process} = [rectangle, 
minimum width=3cm, 
minimum height=1cm, 
text centered, 
text width=3cm, 
draw=black, 
fill=orange!30]

\tikzstyle{decision} = [diamond, 
minimum width=3cm, 
minimum height=1cm, 
text centered, 
draw=black, 
fill=green!30]
\tikzstyle{arrow} = [thick,->,>=stealth]

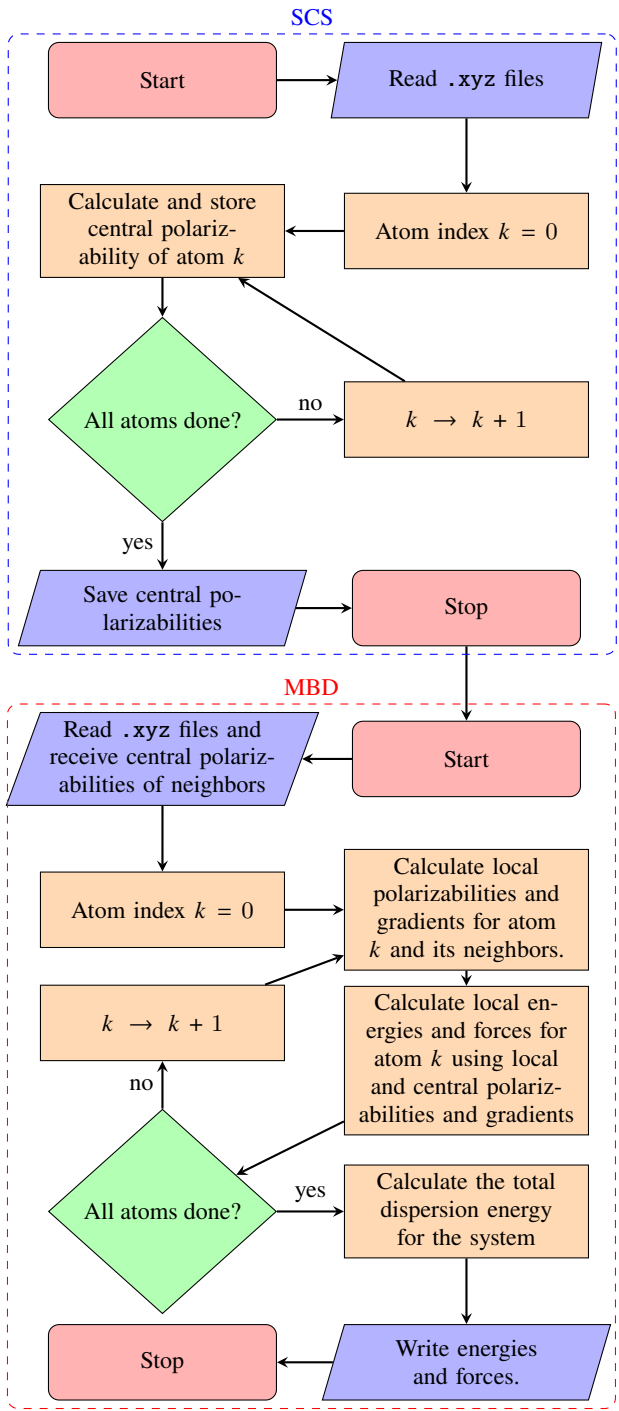
\begin{figure}
\begin{tikzpicture}[node distance=2cm]
\node (start) [startstop] {Start};
\node (read) [io, right of = start, xshift=2cm] {Read \texttt{.xyz} files};
\node (init) [process, below of = read] {Atom index $k = 0$};
\node (central) [process, left of=init, xshift=-2cm] {Calculate and store central polarizability of atom $k$};
\node (done) [decision, below of=central, yshift=-0.5cm] {All atoms done?};
\node (k_iter1) [process, right of=done, xshift=2cm] {$k \rightarrow k+1$};
\node (communicate) [io, below of=done, yshift=-0.5cm] {Save central polarizabilities};
\node (stop_scs) [startstop, right of = communicate, xshift=2cm] {Stop};
\node (start_mbd) [startstop, below of = stop_scs] {Start};
\node (read2) [io, below of = communicate] {Read \texttt{.xyz} files and receive central polarizabilities of neighbors};
\node (start2) [process, below of=read2] {Atom index $k = 0$};
\node (local) [process, right of=start2, xshift=2cm] {Calculate local polarizabilities and gradients for atom $k$ and its neighbors.};
\node (mbd) [process, below of=local] {Calculate local energies
and forces for atom $k$ using local and central polarizabilities and gradients};
\node (k_iter2) [process, below of=start2, yshift=0.5cm] {$k \rightarrow k+1$};
\node (done2) [decision, below of=k_iter2, yshift=-0.5cm] {All atoms done?};
\node (finish) [process, right of=done2, xshift=2cm] {Calculate the total dispersion energy for the system};
\node (output) [io, below of=finish] {Write energies and forces.};
\node (stop) [startstop, left of=output, xshift=-2cm] {Stop};

\draw [arrow] (start) -- (read);
\draw [arrow] (read) -- (init);
\draw [arrow] (init) -- (central);
\draw [arrow] (central) -- (done);
\draw [arrow] (done) -- node[anchor=east] {yes} (communicate);
\draw [arrow] (done) -- node[anchor=south] {no} (k_iter1);
\draw [arrow] (k_iter1) -- (central);
\draw [arrow] (communicate) -- (stop_scs);
\draw [arrow] (stop_scs) -- (start_mbd);
\draw [arrow] (start_mbd) -- (read2);
\draw [arrow] (read2) -- (start2);
\draw [arrow] (start2) -- (local);
\draw [arrow] (local) -- (mbd);
\draw [arrow] (mbd) -- (done2);
\draw [arrow] (done2) -- node[anchor=east] {no} (k_iter2); 
\draw [arrow] (k_iter2) -- (local);
\draw [arrow] (done2) -- node[anchor=south] {yes} (finish);
\draw [arrow] (finish) -- (output);
\draw [arrow] (output) -- (stop);

\node [fit=(start) (read) (init) (central) (done) (k_iter1) (communicate) (stop_scs),draw,dashed,blue,rounded corners,minimum width=8cm,label={[xshift=0.0cm,color=blue]SCS}] {};
\node [fit=(read2) (start2) (local) (mbd) (done2) (k_iter2) (done2) (finish) (output) (stop),draw,dashed,red,rounded corners,minimum width=8cm,label={[xshift=0.0cm,color=red]MBD}] {};

\end{tikzpicture}
\caption{Flowchart of the process for the local approach. The SCS cycle (dashed blue line) can be done separately if one wants to only use the SCS polarizabilities for TS-SCS instead of the full MBD.}
\label{flowchart}
\end{figure}

In the preceding section we formulated the calculation of the non-local MBD energy for a system of $N$ atoms. For periodic (infinite) systems one could sample the Brillouin zone as in the VASP implementation~\cite{buvcko2016many}. However, if the unit cell (or molecule) is too large, this strategy does not help: the matrix related to the problem may become too large to invert (see Eq.~\eqref{polarizabilities}) or diagonalize (see Eq.~\eqref{E_MBD}). We are interested in studying systems where the number of atoms in the unit cell can range from thousands to even millions. For this purpose, it is better to define a local contribution to the energy for each atom within some cutoff to achieve linear scaling. This is also a problem because the total energy defined in Eq.~\eqref{E_MBD} is strictly type-B non-additive~\cite{dobson2014beyond}, whereas a local energy decomposition (as done in \gls{ML} force fields) requires that the local energies be additive.

We will next describe our local approximation in detail. The flow of the process is summarized in Fig.~\ref{flowchart}. We start from the local energies and then proceed to the central and local polarizabilities that are needed to calculate the local polarizability tensor. We first define 
a local atomic neighborhood. If atom $k$ is chosen as the central atom and the local neighborhood (or a ``molecule'') is defined within a certain cutoff radius of atom $k$, one can define matrices $\mathbf{A}_{\text{LR},K}, \mathbf{T}_{\text{LR},K} \in \mathbb{R}^{3N_{\text{neigh},k}\times3N_{\text{neigh},k}}$ that only include the interactions within the full MBD cutoff radius $r_{\text{MBD}}$, between the $N_{\text{neigh},k}$ atoms inside the cutoff sphere. This means that the indices will be a subset of the total list of indices. Using the notation for balls in metric spaces, we will denote the subset of these indices by $K$:
\begin{equation}
K := \{j : \mathbf{r}_j \in \text{B}_{r_{\text{MBD}}} (\mathbf{r}_k) \},
\end{equation}
which we will also use as an index related to the local quantities to differentiate between the different atomic environments (the polarizability of atom $i$ in the $i$-centric local environment might not be equal to the polarizability of $i$ in the $k$-centric environment). 

The total energy that atom $k$ ``sees'' is given by:
\begin{equation}
    E_{\text{MBD},k}^{\text{tot}} = -\int\limits_0^\infty \frac{\text{d}\omega}{2\pi} \text{Tr}\{ \text{ln}(\mathbf{1}-\mathbf{A}_{\text{LR},K}(\omega)\mathbf{T}_{\text{LR},K})\}.
    \label{E_tot}
\end{equation}
If the cutoff spans the entire molecule, it can be easily seen that the original MBD energy is retrieved exactly. This is not necessarily true in periodic systems even if the cutoff is chosen to include the entire unit cell because our method needs to explicitly capture real-space interactions between periodic replicas, which are handled in reciprocal space in the reference method. Let us now expand the logarithm as a power series of matrices (proceeding in the reverse direction from how the logarithm expression was originally obtained~\cite{buvcko2016many}) with certain assumptions:
\begin{equation}
    \text{ln}(\mathbf{1}-\mathbf{A}_{\text{LR},K}\mathbf{T}_{\text{LR},K}) = -\sum\limits_{n=1}^{\infty} \frac{(\mathbf{A}_{\text{LR},K}\mathbf{T}_{\text{LR},K})^n}{n}.
    \label{series-expansion}
\end{equation}
The original derivation assumes that the eigenvalues $\lambda$ of the full matrix $\mathbf{A}_{\text{LR}}\mathbf{T}_{\text{LR}}$ satisfy $|\lambda| < 1$ for convergence. However, in some cases where the atoms are very close to each other, the eigenvalues can be $\lambda < -1$, which causes a polarization catastrophe~\cite{gould2016fractionally}. Indeed, this can happen even for dense carbon structures such as amorphous carbon. VASP handles this problem by introducing eigenvalue remapping and fractionally ionic polarizabilities~\cite{gould2016fractionally}. In this work we will assume that the eigenvalues satisfy $|\lambda| < 1$ because our approach does not 
allow us to use eigenvalue remapping. We elaborate on this issue later in this section.

\begin{figure}
\begin{center}
\begin{tikzpicture}

  \draw(0,0) circle (2.66cm);    
  \draw(0,0) circle (4cm);
  \draw(-0.1,0.35) node {$k$};
  \draw(1,-1) node[anchor=east]{$K_1$};
  \draw(3.5,0) node[anchor=east]{$K_2$};
  \draw[black,fill=red] (0,0) circle (0.2cm);
  \draw[black,fill=blue] (1,1) circle (0.2cm);
  \draw(0.9,1.35) node {$j_1$};
  \draw[black,fill=blue] (-0.3,2.2) circle (0.2cm);
  \draw(-0.05,2.45) node {$j_2$};
  \draw[black,fill=blue] (-1,0.5) circle (0.2cm);
  \draw(-1.1,0.85) node {$j_3$};
  \draw[black,fill=blue] (-2.2,-0.1) circle (0.2cm);
  \draw(-1.95,0.15) node {$j_4$};
  \draw[black,fill=blue] (0,-1.5) circle (0.2cm);
  \draw(-0.1,-1.15) node {$j_5$};
  \draw[black,fill=blue] (2,-0.5) circle (0.2cm);
  \draw(2.25,-0.25) node {$j_6$};
  \draw[black,fill=blue] (2.5,-1.7) circle (0.2cm);
  \draw(2.4,-1.35) node {$j_7$};
  \draw[black,fill=blue] (-3.5,-0.6) circle (0.2cm);
  \draw(-3.6,-0.25) node {$j_8$};
  \draw[black,fill=blue] (3.2,1.6) circle (0.2cm);
  \draw(3.1,1.95) node {$j_9$};
  \draw[black,fill=blue] (-0.4,3.6) circle (0.2cm);
  \draw(-0.7,3.8) node {$j_{10}$};
  \draw[black,fill=blue] (-2.,2.) circle (0.2cm);
  \draw(-2.1,2.35) node {$j_{11}$};
  \draw[black,fill=blue] (-0.9,-2.2) circle (0.2cm);
  \draw(-1,-1.85) node {$j_{12}$};
  \draw[black,fill=blue] (-0.7,-3.3) circle (0.2cm);
  \draw(-0.45,-3.05) node {$j_{13}$};
  \draw[black,fill=blue] (-1.7,-2.7) circle (0.2cm);
  \draw(-1.45,-2.35) node {$j_{14}$};
  \draw[black,fill=blue] (-0.7,-0.5) circle (0.2cm);
  \draw(-0.45,-0.25) node {$j_{15}$};
  \draw[black,fill=blue] (1.3,-2) circle (0.2cm);
  \draw(1.55,-1.75) node {$j_{16}$};
  \draw[black,fill=blue] (-3,0.5) circle (0.2cm);
  \draw(-2.75,0.75) node {$j_{17}$};
  \draw[black,fill=blue] (-0.5,3) circle (0.2cm);
  \draw(-0.25,3.25) node {$j_{18}$};
  \draw[black,fill=blue] (-2.8,2.2) circle (0.2cm);
  \draw(-2.55,2.45) node {$j_{19}$};
  \draw[black,fill=blue] (-3,-1.5) circle (0.2cm);
  \draw(-2.75,-1.25) node {$j_{20}$};
  \draw[black,fill=blue] (0.5,3.2) circle (0.2cm);
  \draw(0.75,3.45) node {$j_{21}$};
  \draw[black,fill=blue] (2.2,2.2) circle (0.2cm);
  \draw(2.45,2.45) node {$j_{22}$};
  \draw[black,fill=blue] (3.5,-0.4) circle (0.2cm);
  \draw(3.75,-0.15) node {$j_{23}$};
  \draw[black,fill=blue] (3.2,0.7) circle (0.2cm);
  \draw(3.45,0.95) node {$j_{24}$};
  \draw[black,fill=blue] (3,-1) circle (0.2cm);
  \draw(3.25,-0.75) node {$j_{25}$};
  \draw[black,fill=blue] (2,-2.8) circle (0.2cm);
  \draw(2.25,-2.55) node {$j_{26}$};
  \draw[black,fill=blue] (-0.9,1.7) circle (0.2cm);
  \draw(-0.65,1.95) node {$j_{27}$};
  \draw[black,fill=blue] (0.4,-3.25) circle (0.2cm);
  \draw(0.3,-2.9) node {$j_{28}$};
  \draw[black,fill=blue] (2.3,2.8) circle (0.2cm);
  \draw(2.2,3.15) node {$j_{29}$};
  \draw[black] (0,0) -- (2.67,0);
  \draw(1.5,0) node[anchor=south east] {$r_{\text{MBD,1}}$};
  \draw[black] (0,0) -- (1,3.87);
  \draw(0.75,3.) node[anchor=north east] {$r_{\text{MBD}}$};
  \draw[dashed] (-0.9,-2.2) circle (1.33cm);
  \draw[dashed] (-0.9,-2.2) -- (0.43,-2.2);
  \draw(0.43,-1.75) node[anchor=north east] {$r_{\text{MBD},2}$};

\end{tikzpicture} 
\end{center}
\caption{Illustrative example of the different MBD cutoff radii for the local energy calculation. Here the central atom is $k$ and the set of indices of atoms in the primary cutoff $r_{\text{MBD},1}$ within the central atom is denoted by $K_1$. Atoms beyond this cutoff are not directly coupled to the central atom but they are coupled to the atoms at the boundary of the cutoff sphere $\text{B}_{r_{\text{MBD,1}}}(\mathbf{r}_k)$ up to the secondary cutoff $r_{\text{MBD},2}$. The set of indices of atoms outside of the primary cutoff $r_{\text{MBD},1}$ but inside the total cutoff $r_{\text{MBD}}$ is denoted by $K_2$. The central atom ``sees'' more atoms than its neighbors because it has the largest effect on the total energy and at the same time the matrix related to the problem becomes sparse. In this example, the central atom is directly coupled to atoms in $K_1: j_1, j_2, j_3, j_4, j_5, j_6, j_{12}, j_{15}, j_{16}$ and $j_{27}$. At the same time, $j_{12}$ is only coupled to atoms $k, j_5, j_{13}$ and $j_{14}$, of which the latter two are in $K_2$ and thus not directly coupled to $k$. Note that even though $k$ is not within $r_{\text{MBD},2}$ of $j_{12}$, it is coupled to that atom because $j_{12}$ is within $r_{\text{MBD},1}$ of $k$ and the couplings have to go both ways to make physical sense: the dipole coupling matrices have to be symmetric.}
\label{mbd-radii}
\end{figure}
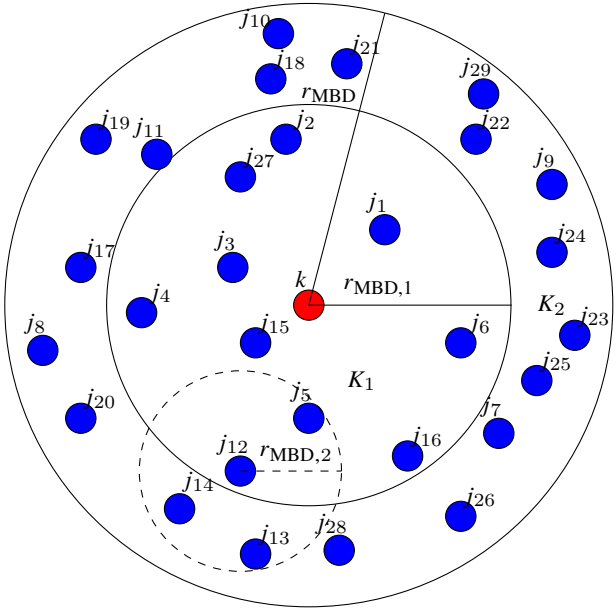

Since $\mathbf{A}_{\text{LR},K}$ is diagonal, it is easy to check that
\begin{equation}
    G_{K,ij}^{\alpha\beta} := [\mathbf{A}_{\text{LR},K} \mathbf{T}_{\text{LR},K}]_{ij}^{\alpha\beta} = \tilde{\alpha}^{\text{iso}}_i (\omega) T_{\text{LR},K,ij}^{\alpha\beta}.
    \label{G_K}
\end{equation}
Next, we define smaller submatrices corresponding to the rows and columns (that is, the different atoms in the cutoff sphere) of $\mathbf{G}_K$. The row matrix corresponding to atom $i$ is given by:
\begin{equation}
    \mathbf{g}_{K,i} = \tilde{\alpha}^{\text{iso}}_i (\omega) \mathbf{T}_{\text{LR},K,i} \in \mathbb{R}^{3\times 3N_{\text{neigh},k}}
\end{equation}
and the column matrix ($\mathbf{T}_{\text{LR},K}$ is symmetric) by:
\begin{multline}
    \mathbf{g}_{K,i}^\dagger = \mathbf{A}_{\text{LR},K} (\mathbf{T}_{\text{LR},K,i})^T \in \mathbb{R}^{3N_{\text{neigh},k} \times 3}, \\ \text{where} \quad [\mathbf{g}_{K,i}^\dagger]_j = \tilde{\alpha}^{\text{iso}}_j (\omega) \mathbf{T}_{\text{LR},K,ji} \in \mathbb{R}^{3\times 3}.
\end{multline}
From this, it follows that the trace of the $n$th power ($n \geq 2$) in Eq.~\eqref{series-expansion} can be written as
\begin{equation}
    \mathrm{Tr}\{(\mathbf{A}_{\text{LR},K}\mathbf{T}_{\text{LR},K})^n\} = \mathrm{Tr}\{\mathbf{G}_K^n\} = \sum\limits_{i \in K} \mathrm{Tr}\{\mathbf{g}_{K,i} \mathbf{G}_K^{n-2} \mathbf{g}_{K,i}^\dagger\}.
\end{equation}
Inserting all of this back into Eq.~\eqref{E_tot} and rearranging the sums we get:
\begin{equation}
    E_{\text{MBD},k}^{\text{tot}} = \sum\limits_{i \in K} \int\limits_0^\infty \frac{\text{d}\omega}{2\pi} \sum\limits_{n=2}^{\infty} \frac{\mathrm{Tr}\{\mathbf{g}_{K,i} \mathbf{G}_K^{n-2} \mathbf{g}_{K,i}^\dagger\}}{n},
    \label{mbd-local-full}
\end{equation}
where the contribution from $n=1$ is zero in all cases (as can be easily checked from the trace). This equation gives an expression for the additive local energies when we only consider the ``subtrace'' corresponding to the central atom $k$, which has the largest contribution (due to the central atom seeing most neighbors):
\begin{equation}
    E_{\text{MBD},k} = \int\limits_0^\infty \frac{\text{d}\omega}{2\pi} \sum\limits_{n=2}^{\infty} \frac{\mathrm{Tr}\{\mathbf{g}_{K,k} \mathbf{G}_K^{n-2} \mathbf{g}_{K,k}^\dagger\}}{n},
\end{equation}
such that the full MBD energy is well approximated by the local contributions:
\begin{equation}
    E_{\text{MBD}} \approx \sum\limits_{k=1}^N E_{\text{MBD},k}.
\end{equation}
We note that the full-\gls{MBD} limit can be systematically retrieved by making the cutoff radius $r_\text{MBD}$ larger; $r_\text{MBD}$ is thus the central convergence parameter controlling the quality of our local decomposition approximation.
Furthermore, the terms with different $n$ correspond to $n$-body interactions. Another cutoff can be introduced to include interactions up to $n_{\text{max}}$-body terms with increasing accuracy:
\begin{equation}
    E_{\text{MBD},k} \approx \int\limits_0^\infty \frac{\text{d}\omega}{2\pi} \sum\limits_{n=2}^{n_{\text{max}}} \frac{\mathrm{Tr}\{\mathbf{g}_{K,k} \mathbf{G}_K^{n-2} \mathbf{g}_{K,k}^\dagger
    \}}{n}.
    \label{mbd-local}
\end{equation}
Note that in the case of two-body interactions the product is computationally inexpensive (because $\text{Tr}\{\mathbf{g}_{K,k} \mathbf{g}_{K,k}^\dagger\}$ is simply a sum of three dot products) but for $n > 2$ it becomes quite expensive for large cutoff radii, including one or more matrix-vector multiplications. To alleviate this problem (and to make forces continuous at the boundary of the cutoff sphere), we split the full MBD cutoff radius $r_{\text{MBD}}$ into two parts: $r_{\text{MBD}} = r_{\text{MBD},1} + r_{\text{MBD},2}$. Here, $r_{\text{MBD},1}$ is the cutoff used to count the neighbors of the central atom and $r_{\text{MBD},2}$ is the cutoff used to count the neighbors of the central atom's neighbors. We assume that $r_{\text{MBD},1} \geq r_{\text{MBD},2}$, so the central atom ``sees'' more atoms than its neighbors. We define two sets of indices related to these cutoff radii:
\begin{flalign}
K_1 &:= \{j : \mathbf{r}_j \in \text{B}_{r_{\text{MBD},1}}(\mathbf{r}_k)\}; \\
K_2 &:= \{j : \mathbf{r}_j \in \text{B}_{r_{\text{MBD}}}(\mathbf{r}_k) \setminus \text{B}_{r_{\text{MBD},1}}(\mathbf{r}_k)\}.
\end{flalign}
These sets and the cutoff radii are schematically shown in Fig~\ref{mbd-radii}. Effectively, the secondary cutoff $r_{\text{MBD},2}$ defines the sparsity of the long-range dipole coupling tensor $\mathbf{T}_{\text{LR},K}$, because only the elements $T_{\text{LR},K,ij}$ ($i \neq k$) for which $r_{ij} < r_{\text{MBD},2}$ are non-zero. For the row corresponding to the central atom $k$, the elements $T_{\text{LR},K,kj}$ for which $r_{kj} < r_{\text{MBD},1}$ are non-zero, remembering that the coupling should be symmetric such that the $T_{\text{LR},K,jk}$ are also non-zero. Defining the cutoff radii and the dipole coupling tensors this way allows one to resort to sparse linear algebra to make the calculations more efficient.

\begin{figure}
\begin{center}
\includegraphics[width=\columnwidth, trim=0.7cm 1cm 0.7cm 1cm]{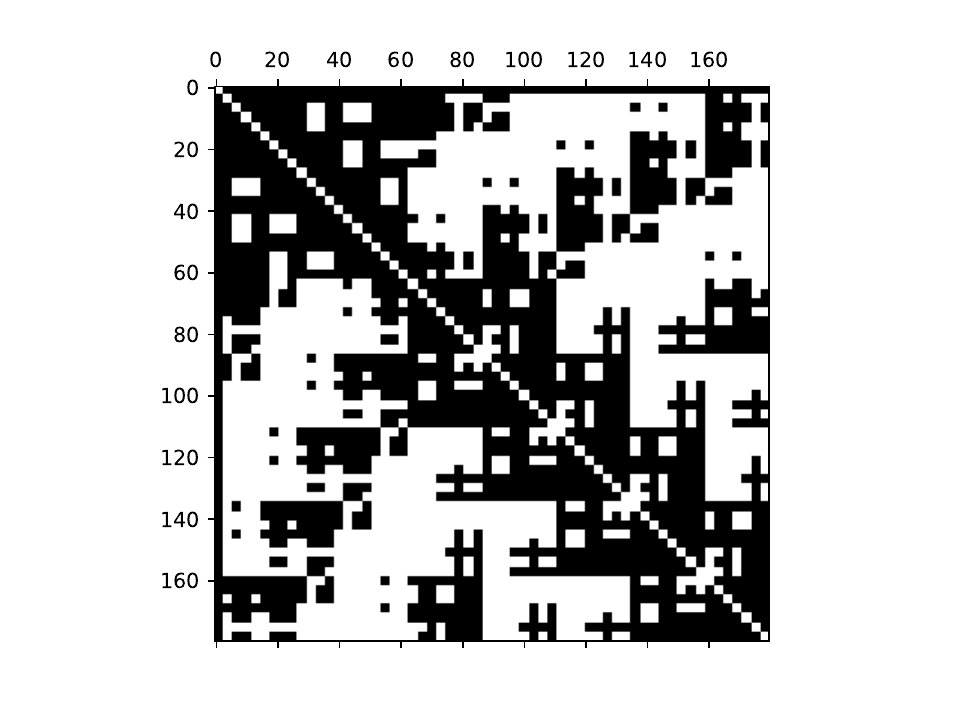}
\end{center}
\caption{Sparsity pattern for a single atom in a C$_{60}$ molecule with $r_{\text{MBD},1} = 10$~Å and $r_{\text{MBD},2} = 5$~Å. The axes correspond to the matrix indices ($180 \times 180$) and the black blocks in the figure are non-zero elements of the matrix $\mathbf{A}_{\text{LR},K}\mathbf{T}_{\text{LR},K}$. The white blocks are zeroes. We have permuted the indices such that the first row and column correspond to the central atom. Because the diameter of the C$_{60}$ molecule is approximately $7$~Å, the central atom ``sees'' all other atoms in the system in this case, and the corresponding rows and columns are full. We have on purpose used a molecule to provide a clear visualization of the sparsity of the matrix but the matrix is not as sparse in this case as it would be for a dense system that would have significantly more matrix elements, such as amorphous carbon.}
\label{sparsity}
\end{figure}

Furthermore, we allow the cutoff radii to be different for the two-body term because it is computationally efficient to calculate (see the $n = 2$ term in Eq.~\eqref{mbd-local}) and has the largest contribution to the energies. The two-body cutoff is controlled by the primary cutoff $r_{\text{2b},1}$ and the secondary cutoff $r_{\text{2b},2}$. For energies, only the primary cutoff is needed and the need for the secondary cutoff will become apparent when we discuss the forces. The sparsity pattern of the matrix $\mathbf{A}_{\text{LR},K}\mathbf{T}_{\text{LR},K}$ for a single atom in a C$_{60}$ molecule using this approach for the cutoff radii is presented in Fig.~\ref{sparsity}.

\begin{figure}
\begin{center}
\includegraphics[width=\columnwidth, trim=0.7cm 1cm 0.7cm 1cm]{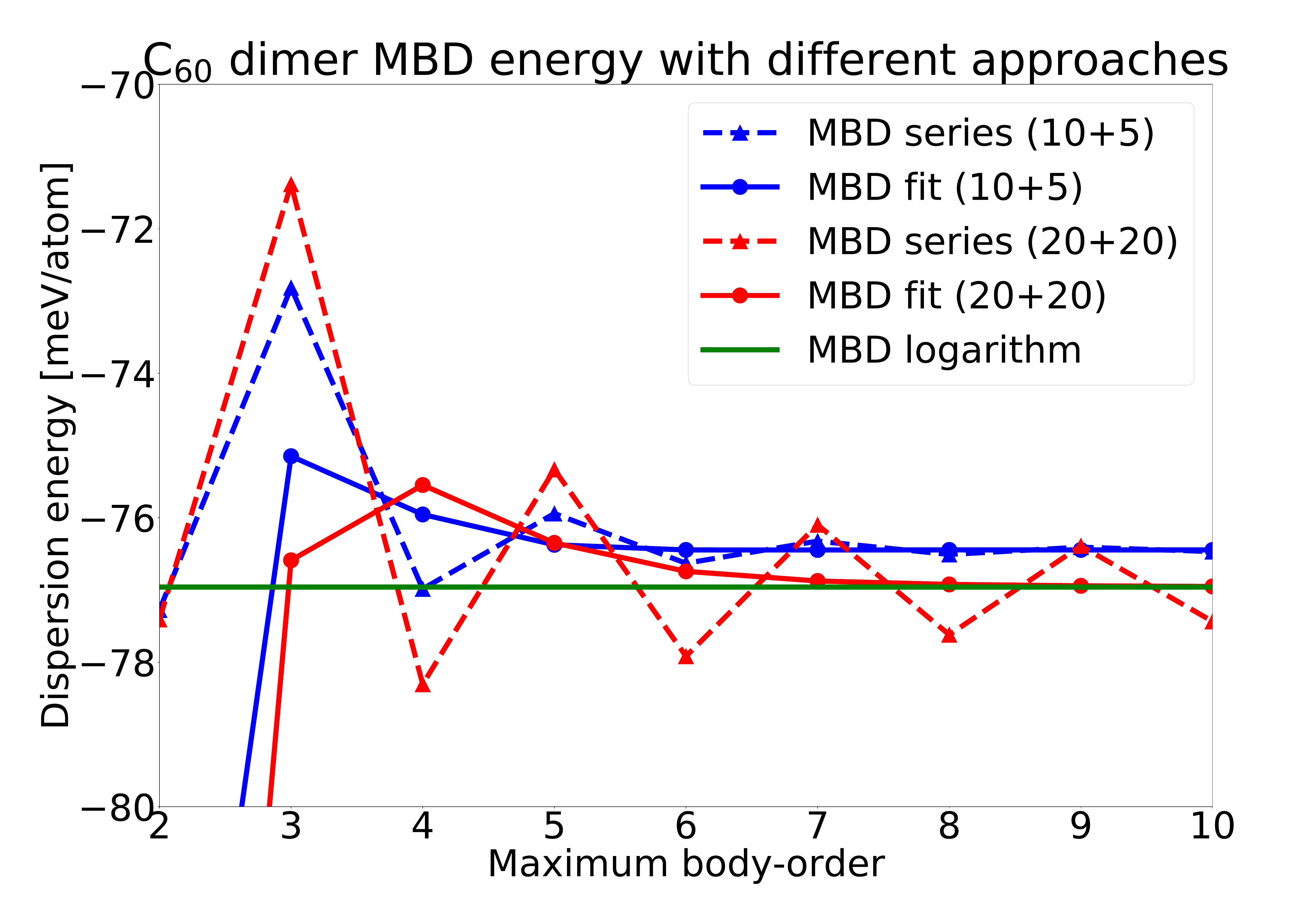}
\end{center}
\caption{Example of the convergence of the series expansion of the matrix logarithm. The figure shows the dispersion energy per atom for a C$_{60}$ dimer calculated with the local approximation described in the text, as a function of the maximum body order of the expansion $n_{\text{max}}$. It can be seen that the oscillating convergence of the series expansion of Eq.~\eqref{mbd-local} (dashed lines) towards the true logarithm value (green) is relatively slow. On the other hand, the polynomial fit of Eq.~\eqref{polynomial-fit} (solid lines) converges swiftly to the limit value as a function of the body order. The numbers in brackets in the legend denote the primary and secondary \gls{MBD} cutoff radii, in \AA{}. Here, 10+5~Å gets close to the logarithm limit and 20+20~Å reaches it because at those values each atom can see the entire dimer.}
\label{convergence}
\end{figure}

There are some technical issues with the expansion of Eq.~\eqref{series-expansion} that need a more detailed discussion. First, the equality is guaranteed to hold only for the eigenvalue spectral radius $\rho(\mathbf{A}_{\text{LR},K} \mathbf{T}_{\text{LR},K}) < 1$~\cite{higham2008functions}, as mentioned earlier. This is not always the case in the systems we have studied. The maximum eigenvalue must not be larger than unity to guarantee convergence. Second, the oscillating convergence of the series can be very slow, as shown in Fig.~\ref{convergence}. Here we assume that the interatomic distances in the system are not too small such that the spectral radius condition is satisfied (to avoid the polarization catastrophe~\cite{gould2016fractionally}) and focus on the convergence of the series with that assumption in mind.

Because of the convergence problem, we decided to provide an alternative solution by fitting a polynomial directly to the desired range of the logarithm values. We estimate the minimum and maximum eigenvalues of $\mathbf{A}_{\text{LR},K}\mathbf{T}_{\text{LR},K}$ using power iteration~\cite{mises1929praktische} or the Lanczos method~\cite{lanczos1950iteration} (after symmetrizing) and denote them by $\lambda_{\text{min},K}$ and $\lambda_{\text{max},K}$ respectively. Then, using least-squares minimization, we fit a polynomial of degree $n_{\text{max}}$ to the scalar function
\begin{equation}
    f(\lambda) = \text{ln}(1-\lambda),
\end{equation}
on the interval $\lambda \in [\lambda_{\text{min},K},\lambda_{\text{max},K}]$. The function is then approximated by:
\begin{equation}
    f(\lambda) \approx \sum\limits_{n=0}^{n_{\text{max}}} c_n \lambda^n,
    \label{diagonal_appr}
\end{equation}
for some coefficients $c_n$. The eigendecomposition of the logarithm of the matrix in Eq.~\eqref{series-expansion} is given by:
\begin{equation}
     \text{ln}(\mathbf{1}-\mathbf{A}_{\text{LR},K}\mathbf{T}_{\text{LR},K}) = \mathbf{V}_K \text{ln}(\mathbf{1} - \boldsymbol{\Lambda}_K)\mathbf{V}_K^{-1},
\end{equation}
where $\boldsymbol{\Lambda}_K$ contains the eigenvalues of $\mathbf{A}_{\text{LR},K}\mathbf{T}_{\text{LR},K}$. Because $\text{ln}(\mathbf{1}-\boldsymbol{\Lambda}_K)$ is a diagonal matrix containing the values $\text{ln}(1-\lambda_{K,i})$ on the diagonal, we can replace these values by their approximations in Eq.~\eqref{diagonal_appr}. Using linear algebra of eigendecompositions then yields:
\begin{equation}
    \text{ln}(\mathbf{1}-\mathbf{A}_{\text{LR},K}\mathbf{T}_{\text{LR},K}) \approx \sum\limits_{n=0}^{n_{\text{max}}} c_n (\mathbf{A}_{\text{LR},K}\mathbf{T}_{\text{LR},K})^n,
    \label{eigendecomp}
\end{equation}
and the local energy in Eq.~\eqref{mbd-local} is given by:
\begin{equation}
    E_{\text{MBD},k} \approx \int\limits_0^\infty \frac{\text{d}\omega}{2\pi} \sum\limits_{n=2}^{n_{\text{max}}} c_n \mathrm{Tr}\{\mathbf{g}_{K,k} \mathbf{G}_K^{n-2} \mathbf{g}_{K,k}^\dagger\}.
    \label{polynomial-fit}
\end{equation}
Here the $n=0$ and $n=1$ terms are always zero, because the trace of $\mathbf{G}_K$ is zero due to zero diagonal and $c_0$ is zero by the requirement that $c_0 \approx f(0) = \text{ln}(1) = 0$. This method avoids the full direct diagonalization of $\mathbf{G}_K$ and instead only requires an iteration for the minimum and maximum eigenvalues and some (sparse) matrix multiplications. The performance of the polynomial fitting compared to the original series expansion can be seen in Fig.~\ref{convergence}.

Furthermore, we note that the equation can be symmetrized, which makes the evaluation even easier. Instead of $\mathbf{G}_K = \mathbf{A}_{\text{LR},K}\mathbf{T}_{\text{LR},K}$, consider $\tilde{\mathbf{G}}_K=\sqrt{\mathbf{A}_{\text{LR},K}}\mathbf{T}_{\text{LR},K}\sqrt{\mathbf{A}_{\text{LR},K}}$. One can check that this conserves the trace
\begin{equation}
    \text{Tr}\{\text{ln}(\mathbf{1}-\mathbf{G}_K)\} = \text{Tr}\{\text{ln}(\mathbf{1}-\tilde{\mathbf{G}}_K)\},
\end{equation}
and makes the matrix $\tilde{\mathbf{G}}_K$ symmetric. The reasoning for the local energy so far remains the same after the symmetrization and the full expression for local energy becomes:
\begin{equation}
    E_{\text{MBD},k} \approx \int\limits_0^\infty \frac{\text{d}\omega}{2\pi} \sum\limits_{n=2}^{n_{\text{max}}} c_n \mathrm{Tr}\{\tilde{\mathbf{g}}_{K,k} \tilde{\mathbf{G}}_K^{n-2} \tilde{\mathbf{g}}_{K,k}^T\},
    \label{E_k}
\end{equation}
where $\tilde{\mathbf{g}}_K^\dagger$ has been replaced by $\tilde{\mathbf{g}}_K^T$ due to the symmetricity of $\tilde{\mathbf{G}}_K$. After symmetrization, it is also possible to apply the Lanczos method instead of power iteration to $\mathbf{1}-\tilde{\mathbf{G}}_K$, which is symmetric and positive definite, to solve for the minimum and maximum eigenvalues of $\tilde{\mathbf{G}}_K$.

Finally, we also redefine the elements of the dipole coupling tensor $T_{\text{LR},K,ij}$ by adding a cutoff function $f_{\text{LR}}(r_{ij})$ to guarantee that the coupling approaches zero smoothly and continuously at the the cutoff:
\begin{equation}
    T_{\text{LR},K,ij}^{\alpha\beta} = f_{\text{LR}}(r_{ij})f(r_{ij},\tilde{S}_{\text{vdW},ij}) T_{ij}^{\alpha\beta}.
\end{equation}
This function is defined as:
\begin{equation}
    f_{\text{LR}}(r_{ij}) = 
    \begin{cases}
        1, & \text{if } r_{ij} < r_{\text{MBD},p} - r_{\text{b}}; \\
        1-3\tilde{r}^2+2\tilde{r}^3, & \text{if } r_{\text{MBD},p} - r_{\text{b}} \leq r_{ij} < r_{\text{MBD},p}; \\
        0, & \text{otherwise},
    \end{cases}
\end{equation}
where $\tilde{r} = ({r_{ij}-r_{\text{MBD},p}+r_{\text{b}}})/{r_{\text{b}}}$ and $p=1$ if $i=k$ or $j=k$ and $p = 2$ otherwise. This means that the central atom has a different cutoff radius than the other atoms, as explained earlier. Here, $r_{\text{b}}$ is the (typically short, $\sim 0.5$~\AA{}) width of the buffer region that guarantees that the coupling smoothly approaches zero at the cutoff to avoid jumps in the energies and forces when atoms enter and leave the cutoff sphere, e.g., during \gls{MD} simulations.

\subsection{Local polarizabilities}

The matrix $\mathbf{A}_{\text{LR},K}$ consists of isotropic \gls{SCS} polarizabilities $\tilde{\alpha}^{\text{iso}}_i (\omega)$ in the local neighborhood of atom $k$. Equation~\eqref{polarizabilities} requires the information from the entire system to solve everything at once and is thus non-local and not amenable to linear scaling. Here we propose a local approximation for solving these polarizabilities. This local approximation only requires information within a small cutoff radius from the central atom to yield polarizabilities that are sufficiently close to the ones given by the full model.

First, we assume that, due to the short-range nature of the dipole coupling (recall that the \gls{SCS} polarizabilities use $\mathbf{T}_{\text{SR}}$ unlike \gls{MBD}, that uses $\mathbf{T}_{\text{LR}}$), for most systems there is no need to include all of the atoms to get a good approximation for the polarizability of the central atom inside a cutoff sphere. We only include neighbors up to the point where the dipole coupling decays to zero due to the damping ($\mathbf{T}_{\text{SR},ij} \approx \mathbf{0}$). Here, we denote the indices inside the SCS cutoff radius $r_{\text{SCS}}$ from the central atom $k$ by set $K_1$ (which is not necessarily the same set we use for the MBD energy calculation discussed earlier). Equation~\eqref{SCS} for the neighborhood $K_1$ then becomes:
\begin{equation}
    \boldsymbol{\tilde{\alpha}}_{K,i} (\omega) \approx \bar{\alpha}_i (\omega)\Big(\mathbf{1}-\sum\limits_{j \in K_1} \mathbf{T}_{\text{SR},K,ij}(\omega) \boldsymbol{\tilde{\alpha}}_{K,j}(\omega)\Big),
    \label{eq:alpha_ki}
\end{equation}
where the definition of the dipole coupling tensor elements $T_{\text{SR},K,ij}$ in the cutoff sphere centered on atom $k$ is given below. However, using Eq.~\eqref{eq:alpha_ki} as is, the atoms at the edge of the cutoff radius would not ``see'' their neighbors and this would result in poor estimates for the boundary polarizabilities, which in turn gives a poor estimate of the central polarizability. To overcome this problem, we introduce uncoupled atoms (i.e., atoms whose polarizabilities are not allowed to change during the \gls{SCS} procedure) beyond the \gls{SCS} cutoff to get more accurate boundary polarizabilities. We choose twice the $r_{\text{SCS}}$ cutoff because this is the furthest the boundary atoms are able to ``see'' due to the damping. Using again the notation for balls in metric spaces, we denote the set of indices in the inner and outer regions by: 
\begin{flalign}
K_1 &:= \{j : \mathbf{r}_j \in \text{B}_{r_{\text{SCS}}}(\mathbf{r}_k)\}, \\
K_2 &:= \{j : \mathbf{r}_j \in \text{B}_{2r_{\text{SCS}}}(\mathbf{r}_k) \setminus \text{B}_{r_{\text{SCS}}}(\mathbf{r}_k)\},
\end{flalign}
respectively, with $K = K_1 \cup K_2$, where $\mathbf{r}_k$ is the position of the central atom. The polarizabilities within $K_1$ (that is, the local neighborhood of atom $k$) are then given by:
\begin{multline}
    \boldsymbol{\tilde{\alpha}}_{K,i} (\omega) \approx \bar{\alpha}_i (\omega)\Big(\mathbf{1}-\sum\limits_{j \in K_1} \mathbf{T}_{\text{SR},K,ij}(\omega) \boldsymbol{\tilde{\alpha}}_{K,j}(\omega) \\ -\sum\limits_{j \in K_2} \mathbf{T}_{\text{SR},K,ij}(\omega) \bar{\alpha}_j(\omega)\Big).
\end{multline}
Here one has to additionally take into account the case where atom $i \in K_1$ (which is always true in the equation above) and atom $j \in K_2$ are very close to each other. The polarizability of $j$ is fixed in this case, which can have an undesirable effect on the polarizability of atom $i$ and through atom $i$ also on other polarizabilities. To prevent this, we introduce an inner buffer function for short range:
\begin{equation}
    f_{\text{inner}}(r_{ij}) = 
    \begin{cases}
        3\Big(\frac{r_{ij}}{r_{\text{inner}}}\Big)^2-2\Big(\frac{r_{ij}}{r_{\text{inner}}}\Big)^3, & \text{if } j \in K_2 \text{ and } r_{ij} < r_{\text{inner}};  \\
        1, & \text{otherwise},
    \end{cases}
\end{equation}
where $r_{\text{inner}}$ is an inner buffer region for the short-range interaction. A sensible value for this parameter is 2~\AA{}, and this is currently hardcoded into our implementation.

Here we briefly mention that in some cases~\cite{gobre2013scaling} the effect of the \gls{SCS} can extend much further and affect the binding properties at a scale much longer than the effective radius of the short-range dipole interaction tensor due to the damping function. The reason for this is that, in the analysis by Gobre \textit{et al}.~\cite{gobre2013scaling}, there is no damping for the \gls{SCS} polarizibility calculation. The range separation of the polarizability and energy calculation was added for the VASP implementation~\cite{buvcko2016many} and, indeed, it seems that VASP would not be able to reproduce the results of Ref.~\onlinecite{gobre2013scaling}, where the extremely long-ranged effect of the \gls{SCS} polarizabilities is seen in the context of the binding energy between a C$_{60}$ molecule and multi-layer graphene. We assume that this effect then has to be implicitly bound to the long-range part of the calculation through range-separation and does not directly show up in the polarizabilities.

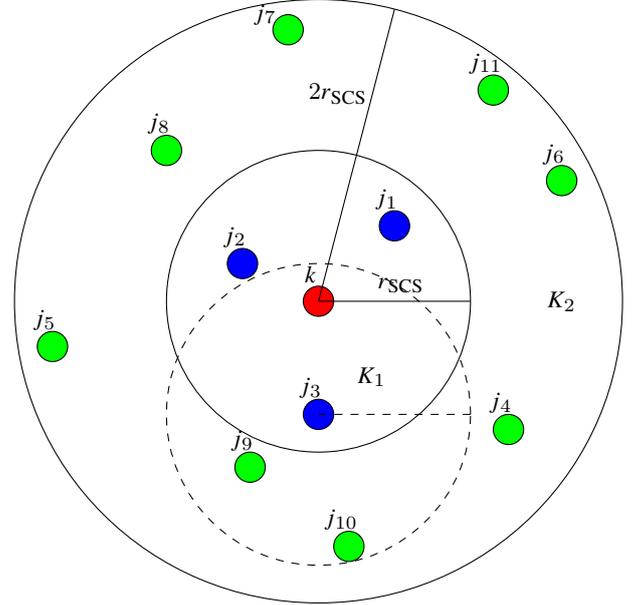
\begin{figure}[t]
\begin{center}
\begin{tikzpicture}
  \draw(0,0) circle (2cm);    
  \draw(0,0) circle (4cm);
  \draw(-0.1,0.35) node {$k$};
  \draw(1,-1) node[anchor=east]{$K_1$};
  \draw(3.5,0) node[anchor=east]{$K_2$};
  \draw[black,fill=red] (0,0) circle (0.2cm);
  \draw[black,fill=blue] (1,1) circle (0.2cm);
  \draw(0.9,1.35) node {$j_1$};
  \draw[black,fill=blue] (-1,0.5) circle (0.2cm);
  \draw(-1.1,0.85) node {$j_2$};
  \draw[black,fill=blue] (0,-1.5) circle (0.2cm);
  \draw(-0.1,-1.15) node {$j_3$};
  \draw[black,fill=green] (2.5,-1.7) circle (0.2cm);
  \draw(2.4,-1.35) node {$j_4$};
  \draw[black,fill=green] (-3.5,-0.6) circle (0.2cm);
  \draw(-3.6,-0.25) node {$j_5$};
  \draw[black,fill=green] (3.2,1.6) circle (0.2cm);
  \draw(3.1,1.95) node {$j_6$};
  \draw[black,fill=green] (-0.4,3.6) circle (0.2cm);
  \draw(-0.7,3.8) node {$j_7$};
  \draw[black,fill=green] (-2.,2.) circle (0.2cm);
  \draw(-2.1,2.35) node {$j_8$};
  \draw[black,fill=green] (-0.9,-2.2) circle (0.2cm);
  \draw(-1,-1.85) node {$j_9$};
  \draw[black,fill=green] (0.4,-3.25) circle (0.2cm);
  \draw(0.3,-2.9) node {$j_{10}$};
  \draw[black,fill=green] (2.3,2.8) circle (0.2cm);
  \draw(2.2,3.15) node {$j_{11}$};
  \draw[black] (0,0) -- (2,0);
  \draw(1.5,0) node[anchor=south east] {$r_{\text{SCS}}$};
  \draw[black] (0,0) -- (1,3.87);
  \draw(0.75,3.) node[anchor=north east] {$2r_{\text{SCS}}$};
  \draw[dashed] (0,-1.5) circle (2cm);
  \draw[dashed] (0,-1.5) -- (2,-1.5);
\end{tikzpicture} 
\end{center}
\caption{Illustration of the calculation of the central (and local) polarizabilities. The inner cutoff sphere is defined by the cutoff radius $r_{\text{SCS}}$ from the central atom $k$ and corresponds in practice to the reach of the dipole coupling tensor $\mathbf{T}_{\text{SR}}$. Atoms that are within this sphere are considered to be in the set $K_1$: $j_1, j_2, j_3 \in K_1$. The distance $2r_{\text{SCS}}$ is the furthest the atoms at the boundary of the inner cutoff sphere can ``see'' due to the range of the dipole coupling tensor. Atoms for which $r_{\text{SCS}} \leq r_{kj} < 2r_{\text{SCS}}$ belong to set $K_2$: $j_4, j_5, \dots , j_{11} \in K_2$. Atoms in $K_1$ contribute to the self-consistent screening, while the atoms in $K_2$ have Tkatchenko-Scheffler polarizabilities and are only coupled to atoms in $K_1$ and thus only enter the right side in Eq.~\eqref{local_pol}. Each atom only couples up to $r_{\text{SCS}}$. For example, the local polarizability of atom $j_3$ is given by: $\tilde{\boldsymbol{\alpha}}_{j_3} = \bar{\alpha}_{j_3}(\mathbf{1} - \mathbf{T}_{\text{SR},j_3k}\tilde{\boldsymbol{\alpha}}_k - \mathbf{T}_{\text{SR},j_3j_9}\bar{\alpha}_{j_9} - \mathbf{T}_{\text{SR},j_3j_{10}} \bar{\alpha}_{j_{10}})$, omitting the frequency dependence for simplicity.}
\label{local_fig}
\end{figure}

Our methodology for the polarizabilities results in a matrix equation of the form (see Eq.~\eqref{polarizabilities}):
\begin{equation}
    \tilde{\mathbf{B}}_K (\omega) \tilde{\boldsymbol{{\alpha}}}_K (\omega) = \mathbf{d}_K (\omega),
    \label{local_pol}
\end{equation}
where $\tilde{\boldsymbol{{\alpha}}}_K (\omega) \in \mathbb{R}^{3 N_{\text{neigh},k} \times 3}$ is a matrix that contains only the SCS polarizabilities of the indices in the set $K_1$,
\begin{equation}
    \tilde{B}_{K,ij}^{\alpha\beta} (\omega) = \frac{1}{\bar{\alpha}_i(\omega)}\delta_{\alpha\beta}\delta_{ij} +  T_{\text{SR},K,ij}^{\alpha\beta}(\omega), \quad i,j \in K_1,
\end{equation}
and
\begin{equation}
    \mathbf{d}_{K,i} (\omega) = \mathbf{1} - \sum\limits_{j \in K_2} \mathbf{T}_{\text{SR},K,ij}(\omega) \bar{\alpha}_j(\omega) \in \mathbb{R}^{3\times 3}.
\end{equation}
This way the uncoupled atoms (in $K_2$) do not contribute to the cost of inverting the matrix $\tilde{\mathbf{B}}_K$ but improve the polarizabilities of the boundary atoms. This process is illustrated in Fig.~\ref{local_fig}.

Similarly to the long-range dipole coupling, we also redefine the dipole coupling elements by adding a cutoff function to guarantee that the elements approach zero smoothly at the cutoff:
\begin{multline}
    T_{\text{SR},K,ij}^{\alpha\beta}(\omega) = f_{\text{inner}}(r_{ij})f_{\text{SR}}(r_{ij})(1-f(r_{ij},S_{\text{vdW},ij})) \\ \times \frac{\partial}{\partial r_i^{\alpha}} \frac{\partial}{\partial r_j^{\beta}} \Bigg(\frac{\text{erf}(r_{ij}/\sigma_{ij}(\omega))}{r_{ij}}\Bigg), 
\end{multline}
where the cutoff function $f_{\text{SR}}(r_{ij})$ is defined as:
\begin{equation}
    f_{\text{SR}}(r_{ij}) = 
    \begin{cases}
        1, & \text{if } r_{ij} < r_{\text{SCS}} - r_{\text{b}}; \\
        1-3\tilde{r}^2+2\tilde{r}^3, & \text{if } r_{\text{SCS}} - r_{\text{b}} \leq r_{ij} < r_{\text{SCS}};  \\
        0, & \text{otherwise},
    \end{cases}
\end{equation}
where $\tilde{r}=({r_{ij}-r_{\text{SCS}}+r_{\text{b}}})/{r_{\text{b}}}$ and $r_{\text{b}}$ again defines the buffer region where the coupling smoothly approaches zero.

\begin{figure}
\begin{center}
\includegraphics[width=\columnwidth, trim=0.7cm 1cm 0.7cm 1cm]{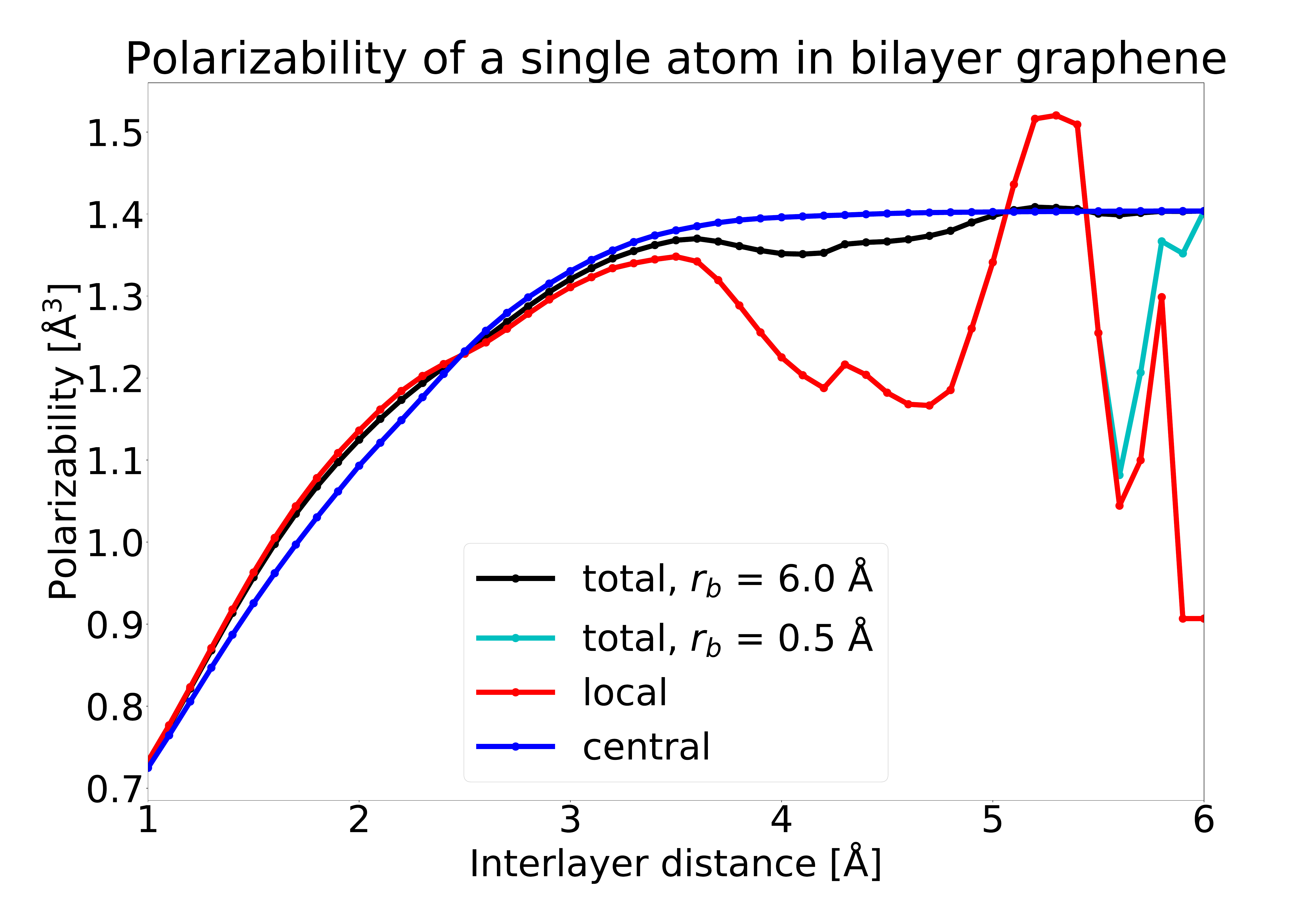}
\end{center}
\caption{The polarizability of a single atom $i$ in bilayer AA stacked graphene as a function of the interlayer distance. The SCS cutoff is $r_{\text{SCS}} = 6.0$ Å. The central (or global) polarizability of atom $i$ here is the polarizability within its own cutoff sphere: $\tilde{\alpha}_{I,i}^{\text{iso}}(0)$. The local polarizability of the same atom given in the cutoff sphere of the central atom $k$ is: $\tilde{\alpha}_{K,i}^{\text{iso}}(0)$. The total polarizability $\tilde{\alpha}_i^{\text{iso}}(0)$ with the buffer width $r_{\rm b} = 6.0$ Å, which equals the cutoff, is given by Eq.~\eqref{full-pol}. If we were to use a cutoff function that uses a buffer width of $r_{\rm b} = 0.5$ Å, that is, local polarizability up to distance of $5.5$ Å and then gradually switching to the central polarizability at $6.0$ Å, the total polarizability would have sharp jumps near the cutoff. This can be seen in the figure where the total polarizability with $r_{\rm b} = 0.5$ Å shares the unstable behavior of the local polarizability close to the cutoff and then suddenly gets the value of the central polarizability at the cutoff. Here one can also see that the central polarizability is converged at about $4$ Å and does not change by further interlayer separation, which means that our approximation for the central polarizability works for bilayer graphene. If one needed to calculate only energies, using the central polarizabilities only would suffice. However, the gradients of the polarizabilities for forces are given by the local part of the polarizabilities.}
\label{AA_pol}
\end{figure}

Because the cutoff $r_{\text{SCS}}$ is usually small as compared to $r_{\text{MBD}}$, solving Eq.~\eqref{local_pol} for the polarizabilities within $r_{\text{SCS}}$ is not very expensive as compared to the local energy calculation that typically uses a much larger cutoff radius. In practice, this equation has to be solved twice, however. We first solve it for each atom in the system to get the central polarizabilities $\tilde{\alpha}_{K,k}$ of the atoms within their own cutoff spheres. These central polarizabilities are then communicated to the local energy calculator where the polarizabilities are solved again to get the local polarizabilities $\tilde{\alpha}_{K,i}$, $i \in K_1$ within the cutoff. This is done because we need the gradients of the local polarizabilities for the force calculation, and communicating all these gradients between the polarizability and the energy calculators for each atom would require too much memory to be feasible. Because of this, the full polarizabilities in the local energy calculation for atom $k$ are given as a range-separated linear combination of the central (global) and local polarizabilities:
\begin{widetext}
\begin{equation}
\tilde{\alpha}_i^{\text{iso}} (\omega) = 
    \begin{cases}
        (1-f_{\text{SCS}}(r_{ik}))\tilde{\alpha}_{K,i}^{\text{iso}}(\omega) + f_{\text{SCS}} (r_{ik}) \tilde{\alpha}_{I,i}^{\text{iso}}(\omega), & \text{if } r_{ik} < r_{\text{SCS}}; \\
        \tilde{\alpha}_{I,i}^{\text{iso}} (\omega), & \text{otherwise}. \\
    \end{cases}
    \label{full-pol}
\end{equation}
\end{widetext}
Intuitively, $\tilde{\alpha}_{I,i}^{\text{iso}}$ can be understood as the polarizability of atom $i$ as seen from its own perspective, whereas $\tilde{\alpha}_{K,i}^{\text{iso}}$ is the polarizability of atom $i$ from the perspective of atom $k$.
Here, $f_{\text{SCS}}(r_{ik})$ is a polynomial function that smoothly switches from 0 to 1:
\begin{equation}
    f_{\text{SCS}}(r_{ik}) = 3\left(\frac{r_{ik}}{r_{\text{SCS}}}\right)^2 - 2\left(\frac{r_{ik}}{r_{\text{SCS}}}\right)^3.
\end{equation}
Unlike in the short-range and long-range damping functions, we use the whole SCS cutoff as the buffer region. This is done to avoid large gradients that the cutoff function would produce because the local polarizability $\tilde{\alpha}_{K,i}^{\text{iso}}(\omega)$ of the atom $i$ at the boundary of the cutoff region $\text{B}_{r_{\text{SCS}}}(\mathbf{r}_k)$ might significantly differ from the polarizability $\tilde{\alpha}_{I,i}^{\text{iso}}(\omega)$ within its own cutoff region $\text{B}_{r_{\text{SCS}}}(\mathbf{r}_i)$ where it is the central atom. The function $f_{\text{SCS}}(r_{ik})$ always depends on the position of the central atom $k$ and appears on each element of the dipole coupling tensor. The gradients of such elements can have a large unwanted effect on the forces if the buffer width is too small because the local polarizabilities can be unstable near the boundary. This problem is illustrated in Fig~\ref{AA_pol}.

\subsection{Frequency dependence}

Because the polarizabilities depend on the frequency $\omega$ and the frequency dependence in Eq.~\eqref{polarizabilities} is not separable, one would have to calculate the SCS polarizabilities for a range of values. We found that the SCS polarizabilities follow roughly the same Lorentzian shape that is used in the polarizabilities of the uncoupled atoms (see Eq.~\eqref{bar-alpha}) but with a different characteristic frequency constant. Based on this, we can approximate:
\begin{equation}
    \tilde{\alpha}^{\text{iso}}_i (\omega) \approx \frac{\tilde{\alpha}^{\text{iso}}_i (0)}{1+(\frac{\omega}{\tilde{\omega}_i} )^2},
    \label{alpha_SCS_freq}
\end{equation}
where $\tilde{\omega}_i$ can be solved for by calculating $\tilde{\alpha}^{\text{iso}}_i (\omega)$ for two frequency values. We use the values $\omega = 0$ and some reference value $\omega = \omega_{\text{ref}}$ such that
\begin{equation}
    \tilde{\omega}_i = \frac{\omega_{\text{ref}}}{\sqrt{\frac{\tilde{\alpha}^{\text{iso}}_i(0)}{\tilde{\alpha}^{\text{iso}}_i(\omega_{\text{ref}})}-1}}.
    \label{char-freq}
\end{equation}
The choice of $\omega_{\text{ref}}$ can slightly affect the value of $\tilde{\omega}_i$ but we have found that the best value for $\omega_{\text{ref}}$ mostly depends on the chemical elements and treat it as an empirical parameter. If one wants to fit the parameter more accurately, more frequency points can be calculated and then a least-squares fit can be applied.

Given the assumed frequency dependence of the SCS polarizabilities in Eq.~\eqref{alpha_SCS_freq}, the separated frequency dependence of the matrix element $\tilde{G}_{K,ij}$ in the local energy calculation is given by:
\begin{align}
    \tilde{G}_{K,ij}(\omega) & = \sqrt{\tilde{\alpha}_i^{\text{iso}}(\omega)\tilde{\alpha}_j^{\text{iso}}(\omega)} T_{\text{LR},K,ij} \nonumber \\ & = \sqrt{\frac{1}{1+(\frac{\omega}{\tilde{\omega}_i} )^2}\frac{1}{1+(\frac{\omega}{\tilde{\omega}_j} )^2}} \tilde{G}_{K,ij}(0).
    \label{G_frequency}
\end{align}
Because the frequency $\omega$ \textit{cannot} be separated from the characteristic frequencies such that we would have:
\begin{equation}
    \tilde{G}_{K,ij}(\omega) \stackrel{!}{=} s(\omega) t(\omega_i,\omega_j) \tilde{G}_{K,ij}(0)
\end{equation}
for some scalar functions $s: \mathbb{R} \rightarrow \mathbb{R}$ and $t: \mathbb{R}\times \mathbb{R} \rightarrow \mathbb{R}$, it also does not separate for the elements $[\tilde{\mathbf{G}}_K(\omega)^n]_{ij}$. Calculating this multiple times for different frequency values to obtain the integrand in Eq.~\eqref{E_k} is expensive. This is why we once again use fitting to make the computation more efficient.
We calculate the integrand in Eq.~\eqref{E_k},
\begin{equation}
I(\omega)=\sum\limits_{n=2}^{n_{\text{max}}} c_n \text{Tr} \{\tilde{\mathbf{g}}_{K,k}(\omega)\tilde{\mathbf{G}}_K(\omega)^{n-2}\tilde{\mathbf{g}}_{K,k}(\omega)^T\},
\label{integrand}
\end{equation}
for at least $n_{\text{max}}+1$ values and then use non-negative least-squares fitting~\cite{lawson1995solving} to obtain the approximate frequency dependence of the integrand:
\begin{equation}
    I(\omega) \approx \frac{a}{1+\sum\limits_{n=1}^{n_{\text{max}}} b_n \omega^{2n}} I(0) = \phi(\omega) I(0),
\end{equation}
for some coefficients $a$ and $b_n$. The equation above defines the function $\phi(\omega)$. This function has the same approximate frequency dependence as the $n_{\text{max}}$ term in Eq.~\eqref{integrand} but it approximates the average frequency dependence of the integrand (which is a scalar) as a whole instead. Non-negative least squares has to be used for the denominator because of the possible singularities at
\begin{equation}
    1+\sum\limits_{n=1}^{n_{\text{max}}} b_n \omega^{2n} = 0,
\end{equation}
which are possible if we allow for negative $b_n$. Furthermore, by looking at Eq.~\eqref{G_frequency}, we infer that these coefficients should be positive if the characteristic frequencies $\tilde{\omega}_i$ are positive for all $i$. Using this approach for the frequency dependence of the integrand, we can obtain very accurate values for the integral by calculating the integrand for just a few frequency values. The full expression for the energies with this approach for the frequency is given by:
\begin{equation}
    E_{\text{MBD},k} \approx \int\limits_0^\infty \frac{\text{d}\omega}{2\pi} \phi(\omega)\sum\limits_{n=2}^{n_{\text{max}}} c_n \mathrm{Tr}\{\tilde{\mathbf{g}}_{K,k}(0) \tilde{\mathbf{G}}_K^{n-2} (0)\tilde{\mathbf{g}}_{K,k}(0)^T\}.
    \label{E_MBD_k}
\end{equation}

\subsection{Dispersion forces}

As mentioned in the Introduction, it is important for efficient \gls{MD}
to have analytic expressions for the forces. Their derivation is rather straightforward but cumbersome due to the complex expressions for the polarizabilities and the energies. The forces acting on atom $k$ are obtained as the negative gradient of the total energy the atom $k$ ``sees'':
\begin{equation}
    f_k^\gamma = -\frac{\partial E_{\text{MBD}}^{k,\text{tot}}}{\partial r_k^\gamma}.
\end{equation}
Here $\gamma$ denotes the Cartesian component. This is mostly a straightforward differentiation of the following equation:
\begin{equation}
    E_{\text{MBD},k}^{\text{tot}} \approx \int\limits_0^\infty \frac{\text{d}\omega}{2\pi} \sum\limits_{n=2}^{n_{\text{max}}} c_n \mathrm{Tr}\{ \tilde{\mathbf{G}}_K(\omega)^n\}.
    \label{E_k_tot}
\end{equation}
The complicated full expression of the derivative is omitted here but its explicit form can be consulted from our reference Fortran implementation in the TurboGAP code~\cite{ref_turbogap}. Instead, we focus on two steps that require some additional work. These are the derivatives of the SCS polarizabilities $\boldsymbol{\tilde{\alpha}}_i$ and the Hirshfeld volumes $\nu_i$.

First, we examine the derivatives of the Hirshfeld volumes. As non-variational functionals of the electron density, the Hellmann-Feynman theorem does not apply and, in principle, analytical gradients of the $\nu_i$ are not available. However, if an analytically differentiable \gls{ML} model for $\nu_i$ exists, these gradients can be easily computed and can indeed provide more accurate dispersion forces than those directly available from a \gls{DFT} calculation~\cite{muhli2021machine}.
Thus, within our \gls{ML} framework for $\nu_i$ prediction~\cite{muhli2021machine}, the derivatives of the Hirshfeld volumes are given by differentiating Eq.~\eqref{kernel-expression}, where the Hirshfeld volume is given by an expression involving \gls{SOAP} descriptors:
\begin{equation}
    \frac{\partial\nu_i}{\partial r_k^\gamma} = \sum\limits_{s\in S} \alpha_s \zeta (\mathbf{q}_s \mathbf{q}_i^T)^{\zeta-1} \mathbf{q}_s\frac{\partial \mathbf{q}_i^T}{\partial r_k^\gamma}.
\end{equation}
These derivatives of the \gls{SOAP} descriptors ${\partial \mathbf{q}_i}/{ \partial r_k^\gamma}$ are also calculated within the \gls{ML} model.

The second point is the differentiation of the SCS polarizabilities. This is done by first differentiating the equation (see Eq.~\eqref{local_pol}, we omit the frequency dependence for simplicity):
\begin{equation}
    \mathbf{\tilde{B}}_K \boldsymbol{\tilde{\alpha}}_K = \mathbf{d}_K,
\end{equation}
with respect to $r_k^\gamma$, which gives
\begin{equation}
    \frac{\partial \mathbf{\tilde{B}}_K}{\partial r_k^\gamma} \boldsymbol{\tilde{\alpha}}_K + \mathbf{\tilde{B}}_K \frac{\partial \boldsymbol{\tilde{\alpha}}_K}{\partial r_k^\gamma} = \frac{\partial \mathbf{d}_K}{\partial r_k^\gamma},
\end{equation}
using the product rule. After trivial linear algebra we get:
\begin{equation}
    \frac{\partial \boldsymbol{\tilde{\alpha}}_K}{\partial r_k^\gamma} = \mathbf{\tilde{B}}_K^{-1} \Big(\frac{\partial \mathbf{d}_K}{\partial r_k^\gamma}-\frac{\partial \mathbf{\tilde{B}}_K}{\partial r_k^\gamma} \boldsymbol{\tilde{\alpha}}_K \Big).
\end{equation}
As opposed to the central polarizability calculation, we need all of the elements of the matrix ${\partial \boldsymbol{\tilde{\alpha}}_K}/{\partial r_k^\gamma} \in 3N_{\text{neigh},k}\times 3$ to accurately calculate the forces, as mentioned earlier. The gradient of the isotropic polarizability is also given by one third of the trace of the corresponding block matrix:
\begin{equation}
     \frac{\partial \tilde{\alpha}^{\text{iso}}_{K,i}}{\partial r_k^\gamma} = \frac{1}{3} \text{Tr}\left\{ \frac{\partial \boldsymbol{\tilde{\alpha}}_{K,i}}{\partial r_k^\gamma}\right\}.
\end{equation}
The differentiation of the range-separated polarizability in Eq.~\eqref{full-pol} is then given by:
\begin{widetext}
\begin{equation}
\frac{\partial\tilde{\alpha}_i^{\text{iso}}}{\partial r_k^\gamma} \approx 
    \begin{cases}
        \frac{\partial f_{\text{SCS}}(r_{ik})}{\partial r_k^\gamma}(\tilde{\alpha}_{I,i}^{\text{iso}} - \tilde{\alpha}_{K,i}^{\text{iso}}) + (1-f_{\text{SCS}}(r_{ik})) \frac{\partial \tilde{\alpha}^{\text{iso}}_{K,i}}{\partial r_k^\gamma}, & \text{if } r_{ik} < r_{\text{SCS}}; \\
        0, & \text{otherwise}, \\
    \end{cases}
    \label{full-pol-grad}
\end{equation}
\end{widetext}
where we have used the approximation ${\partial \tilde{\alpha}_{I,i}}/{\partial r_k^\gamma} = 0$, because we do not communicate the derivatives from the calculation of the central polarizabilities to the local energy calculator. We thus assume that when an atom is moved the change in the surrounding polarizabilites is mostly local due to the short-range damping of the SCS polarizabilities. We use the derivatives of the isotropic polarizabilities to calculate the derivatives of the characteristic frequencies in Eq.~\eqref{char-freq}.

The derivative of the trace in Eq.~\eqref{E_k_tot} (omitting frequency dependence for now) is given by (using the cyclicity of the trace in the second equality):
\begin{align}
    \frac{\partial}{\partial r_k^\gamma} \mathrm{Tr} (\tilde{\mathbf{G}}_K^n) & = \sum\limits_{j=1}^n \mathrm{Tr}(\tilde{\mathbf{G}}_K^{j-1} \frac{\partial \tilde{\mathbf{G}}_K}{\partial r_k^\gamma} \tilde{\mathbf{G}}_K^{n-j}) \nonumber \\ & = n\mathrm{Tr}(\tilde{\mathbf{G}}_K^{n-1} \frac{\partial \tilde{\mathbf{G}}_K}{\partial r_k^\gamma}),
\end{align}
which gives the MBD forces as:
\begin{align}
    f_{\text{MBD},k}^\gamma & = -\frac{\partial E_{\text{MBD}}^{k,\text{tot}}}{\partial r_k^\gamma} \nonumber \\ & \approx -\int\limits_0^\infty \frac{\mathrm{d}\omega}{2\pi} \sum\limits_{n=2}^{n_{\text{max}}} c_n n \mathrm{Tr} \Big(\tilde{\mathbf{G}}_K^{n-1} \frac{\partial \tilde{\mathbf{G}}_K}{\partial r_k^\gamma}\Big).
    \label{mbd-forces}
\end{align}

It is worth noting that the derivative is zero for some elements of the matrix $\tilde{\mathbf{G}}_K$. The cutoff radius $r_{\text{forces}}$ needed for the matrix to calculate the derivative (the elements are zero by construction beyond this cutoff) is given by:
\begin{equation}
    r_{\text{forces}} = \text{max}\{r_{\text{MBD},1}+r_{\text{MBD},2},r_{\text{SCS}}+2r_{\text{MBD},2}\},
\end{equation}
where $r_{\text{MBD},1} \geq r_{\text{SCS}}$ is assumed. Note that for the energy calculation the cutoff is simply $r_{\text{MBD},1}+r_{\text{MBD,2}}$. Due to the total energy involved in Eq.~\eqref{mbd-forces} above, the cost of calculating the forces is still significantly higher than the calculation of local energies and becomes very expensive for dense systems such as amorphous carbon. However, there is a further approximation that brings the cost of the force calculation down to the same level as the local energy calculation. We can approximate the gradients of the polarizabilities to zero for $n \geq 3$ terms when they are not coupled to the central atom:
\begin{equation}
    \frac{\partial\tilde{G}_{K,ij}}{\partial r_k^\gamma} \approx \sqrt{\tilde{\alpha}^{\text{iso}}_i\tilde{\alpha}^{\text{iso}}_j} \frac{\partial T_{\text{LR},K,ij}}{\partial r_k^\gamma}, \ \text{for } i \neq k \text{ and } j \neq k,
\end{equation}
and
\begin{equation}
    \frac{\partial \tilde{S}_{\text{vdW},ij}}{\partial r_k^\gamma} \approx 0, \ \text{for } i \neq k \text{ and } j \neq k.
\end{equation}
This kind of approximation for the forces has been done before, for example in Ref.~\onlinecite{ambrosetti2014long}, and we employ it again here to mitigate the high computational cost of the forces for dense systems that comes from our local approach. This way the only non-zero rows and columns of $\partial\tilde{\mathbf{G}}_K/\partial r_k^\gamma$ are the ones where the index matches $k$:
\begin{equation}
    \frac{\partial\tilde{\mathbf{G}}_K}{\partial r_k^\gamma} \approx \frac{\partial \tilde{\mathbf{G}}_{K,k}}{\partial r_k^\gamma} + \frac{\partial \tilde{\mathbf{G}}_{K,k}^T}{\partial r_k^\gamma},
\end{equation}
where $\tilde{\mathbf{G}}_{K,k}$ represents the matrix $\tilde{\mathbf{G}}_K$ with only the rows $\tilde{\mathbf{g}}_{K,k} \in \mathbb{R}^{3 \times 3N_{\text{neigh},k}}$ as non-zero (note that the block-diagonal of $\tilde{\mathbf{G}}_K$ is also zero). Because of the cyclicity of the trace, the symmetricity of the matrix, and the zero rows of the matrix, Eq.~\eqref{mbd-forces} can be written as (note that we separate the $n = 2$ term for further analysis):
\begin{multline}
    f_{\text{MBD},k}^\gamma \approx -\int\limits_0^\infty \frac{\mathrm{d}\omega}{2\pi} 2c_2 \mathrm{Tr}\Big(\tilde{\mathbf{G}}_K \frac{\partial\tilde{\mathbf{G}}_K}{\partial r_k^\gamma}\Big) \\ +\int\limits_0^\infty \frac{\mathrm{d}\omega}{\pi}\sum\limits_{n=3}^{n_{\text{max}}} c_n n \mathrm{Tr} \Big(\tilde{\mathbf{g}}_{K,k}\tilde{\mathbf{G}}_K^{n-2} \frac{\partial \tilde{\mathbf{g}}_{K,k}^T}{\partial r_k^\gamma}\Big),
    \label{cent-appr-step}
\end{multline}
where the cost of evaluating the sum is now equal to the cost of the local energy calculation with the aforementioned approximation: multiplication of a dense $3\times 3N_{\text{neigh},k}$ matrix with a sparse $3N_{\text{neigh},k}\times 3N_{\text{neigh},k}$ matrix $n_{\text{max}}-2$ times and then taking the dot product with the resulting three vectors of length $3N_{\text{neigh},k}$. Because the matrices $\tilde{\mathbf{g}}_{K,k}\tilde{\mathbf{G}}_{K}^{n-2}$ also appear in the energy calculation, these can be performed simultaneously.

This approximation is done only for the terms where $n \geq 3$. The $n = 2$ term which we have separated in Eq.~\eqref{cent-appr-step} can be calculated efficiently without resorting to matrix multiplication:
\begin{equation}
    \mathrm{Tr}\Big(\tilde{\mathbf{G}}_K \frac{\partial\tilde{\mathbf{G}}_K}{\partial r_k^\gamma}\Big) = \sum\limits_{i,j}\sum\limits_{\alpha,\beta}\Big[\tilde{\mathbf{G}}_K \odot \frac{\partial\tilde{\mathbf{G}}^T_K}{\partial r_k^\gamma}\Big]_{ij}^{\alpha\beta},
\end{equation}
where $\odot$ denotes the Hadamard product (element-wise product) of the two matrices and the trace becomes a grand sum (sum of all elements) of the resulting matrix. The total dispersion forces are then given by:
\begin{multline}
        f_{\text{MBD},k}^\gamma \approx -\int\limits_0^\infty \frac{\mathrm{d}\omega}{\pi} \Big[c_2 \sum\limits_{i,j}\sum\limits_{\alpha,\beta}\Big[\tilde{\mathbf{G}}_K \odot \frac{\partial\tilde{\mathbf{G}}_K}{\partial r_k^\gamma}\Big]_{ij}^{\alpha\beta}+\\\sum\limits_{n=3}^{n_{\text{max}}} c_n n \mathrm{Tr} \Big(\tilde{\mathbf{g}}_{K,k}\tilde{\mathbf{G}}_K^{n-2} \frac{\partial \tilde{\mathbf{g}}_{K,k}^T}{\partial r_k^\gamma}\Big)\Big],
    \label{cent-appr}
\end{multline}
where we have dropped the transpose in the two-body term because of the symmetricity of the matrix. As we mentioned in the discussion of local energies, a separate pair of cutoff radii $r_{\text{2b},1}$ and $r_{\text{2b},2}$ can be used for the two-body term. Here the significance of the secondary two-body cutoff becomes apparent: it defines the sparsity of the two-body term in the force calculation (i.e., how many atoms the neighbors ``see'' when the central atom is moving). The accuracy of this approximation is discussed in the next section.

The force integrand also depends on the frequency, a fact we have ignored in the derivation thus far. The frequency dependence of the force integrand is handled in a similar fashion to the local energy integrand; however, our tests indicate that the product of Lorentzian-type functions may not be a good approximation for the resulting integrand, with the original function potentially changing sign but the fitted one constrained to being positive by the assumed functional form. Instead, the force integrand is calculated for some frequency values and then a function of the following type is fitted from the results:
\begin{equation}
    \theta(\omega) = \frac{\sum\limits_{m=0}^M u_m (-\omega)^m}{1+\sum\limits_{k=1}^K v_k \omega^{2n}},
\end{equation}
for some integers $M \leq n_{\text{max}}$ and $K \leq n_{\text{max}}$. At least $M+K+1$ initial integrand values at different frequencies are needed to solve for the coefficients. We use non-negative least squares~\cite{lawson1995solving} to obtain the coefficients $u_m$ and $v_k$. This approach lets the approximating function change sign, which is what often happens with the force integrand, without singularities. This function can be seen as a Padé approximant~\cite{pade1892representation,baker1981morris} of order $[M/K]$ for the integrand but with non-negative coefficients. The performance of this approximation is also explored in the next section. With this fitting scheme, the forces in Eq.~\eqref{mbd-forces} can be written as:
\begin{multline}
    f_{\text{MBD},k}^\gamma \approx -\int\limits_0^\infty \frac{\mathrm{d}\omega}{2\pi} \theta(\omega)\\ \times \sum\limits_{n=2}^{n_{\text{max}}} c_n n \mathrm{Tr} \Big(\tilde{\mathbf{G}}_K(0)^{n-1} \frac{\partial \tilde{\mathbf{G}}_K(0)}{\partial r_k^\gamma}\Big),
    \label{mbd-forces-sym}
\end{multline}
or for the central atom approximation in Eq.~\eqref{cent-appr} equivalently as:
\begin{multline}
    f_{\text{MBD},k}^\gamma \approx -\int\limits_0^\infty \frac{\mathrm{d}\omega}{\pi} \theta(\omega)\Big[c_2 \sum\limits_{i,j}\sum\limits_{\alpha,\beta}\Big[\tilde{\mathbf{G}}_K(0) \odot \frac{\partial\tilde{\mathbf{G}}_K(0)}{\partial r_k^\gamma}\Big]_{ij}^{\alpha\beta}+\\ \times \sum\limits_{n=3}^{n_{\text{max}}} c_n n \mathrm{Tr} \Big(\tilde{\mathbf{g}}_{K,k}(0)\tilde{\mathbf{G}}_K(0)^{n-2} \frac{\partial \tilde{\mathbf{g}}_{K,k}(0)^T}{\partial r_k^\gamma}\Big)\Big].
\end{multline}

\begin{figure*}
\begin{tabular}{c c}
\includegraphics[width=\columnwidth, trim=0.7cm 1cm 0.7cm 0cm]{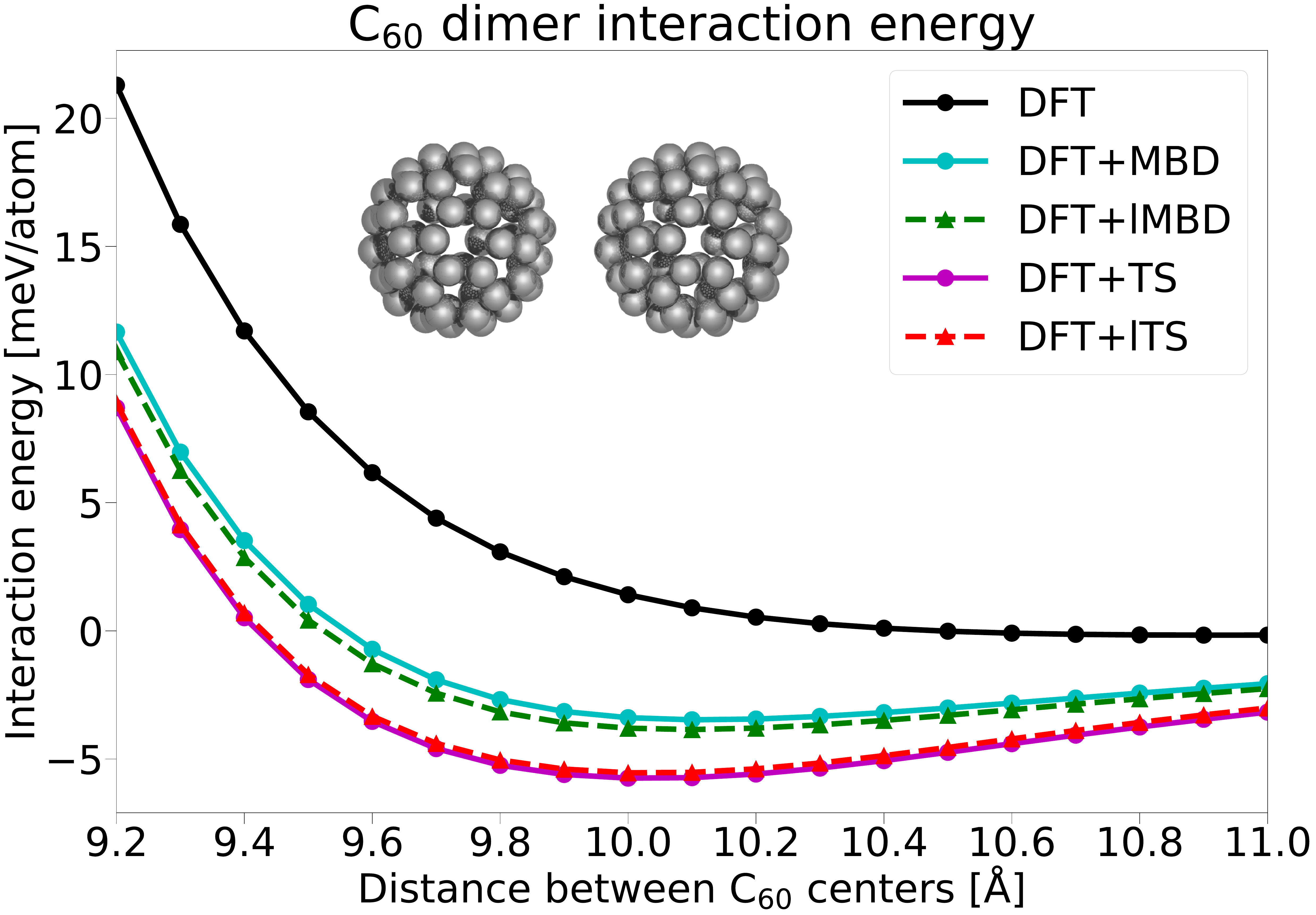}
&
\includegraphics[width=\columnwidth, trim=0.7cm 1cm 0.7cm 0cm]{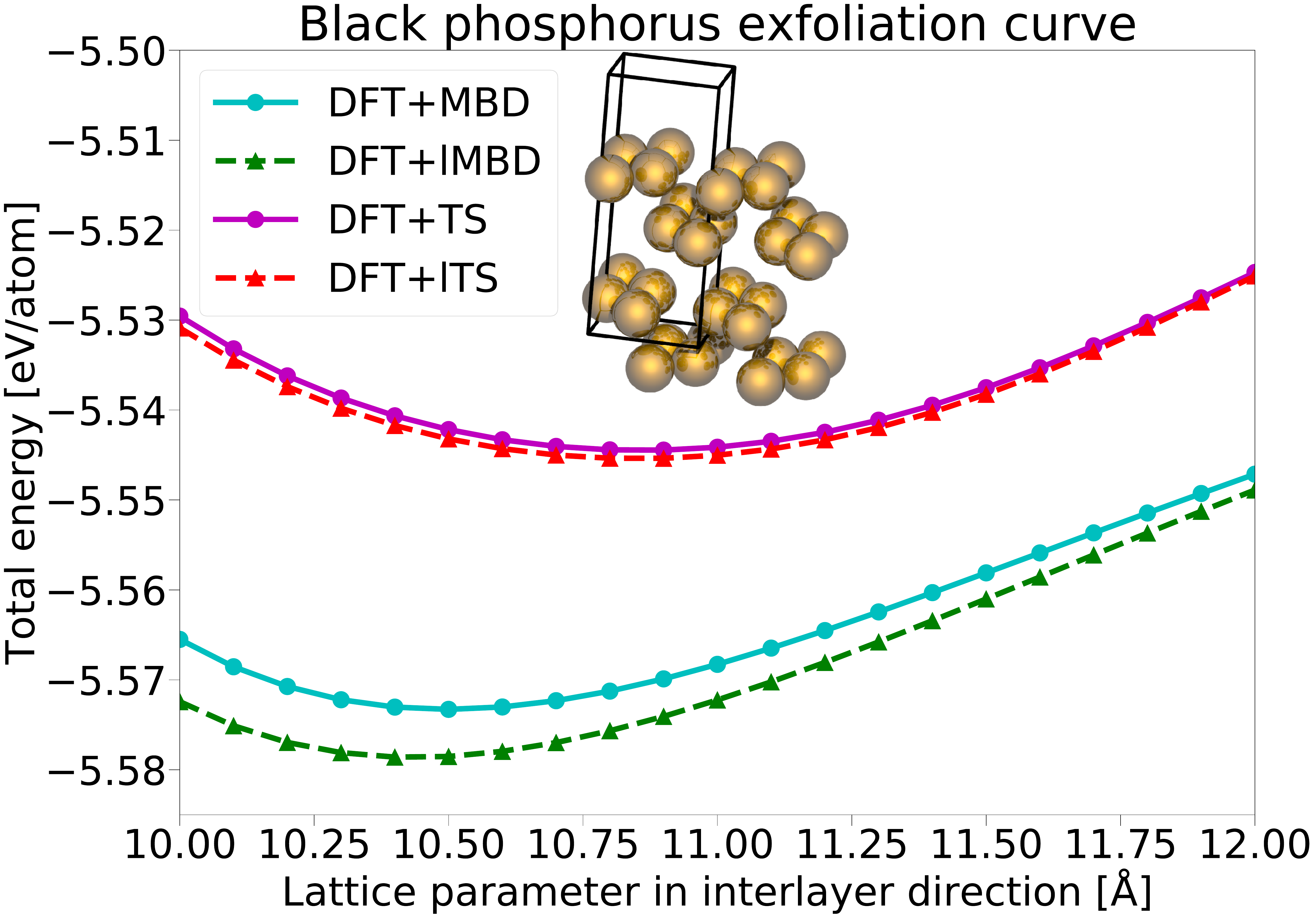}
\\
(a)
&
(b)
\\
\includegraphics[width=\columnwidth, trim=0.7cm 1cm 0.7cm 0cm]{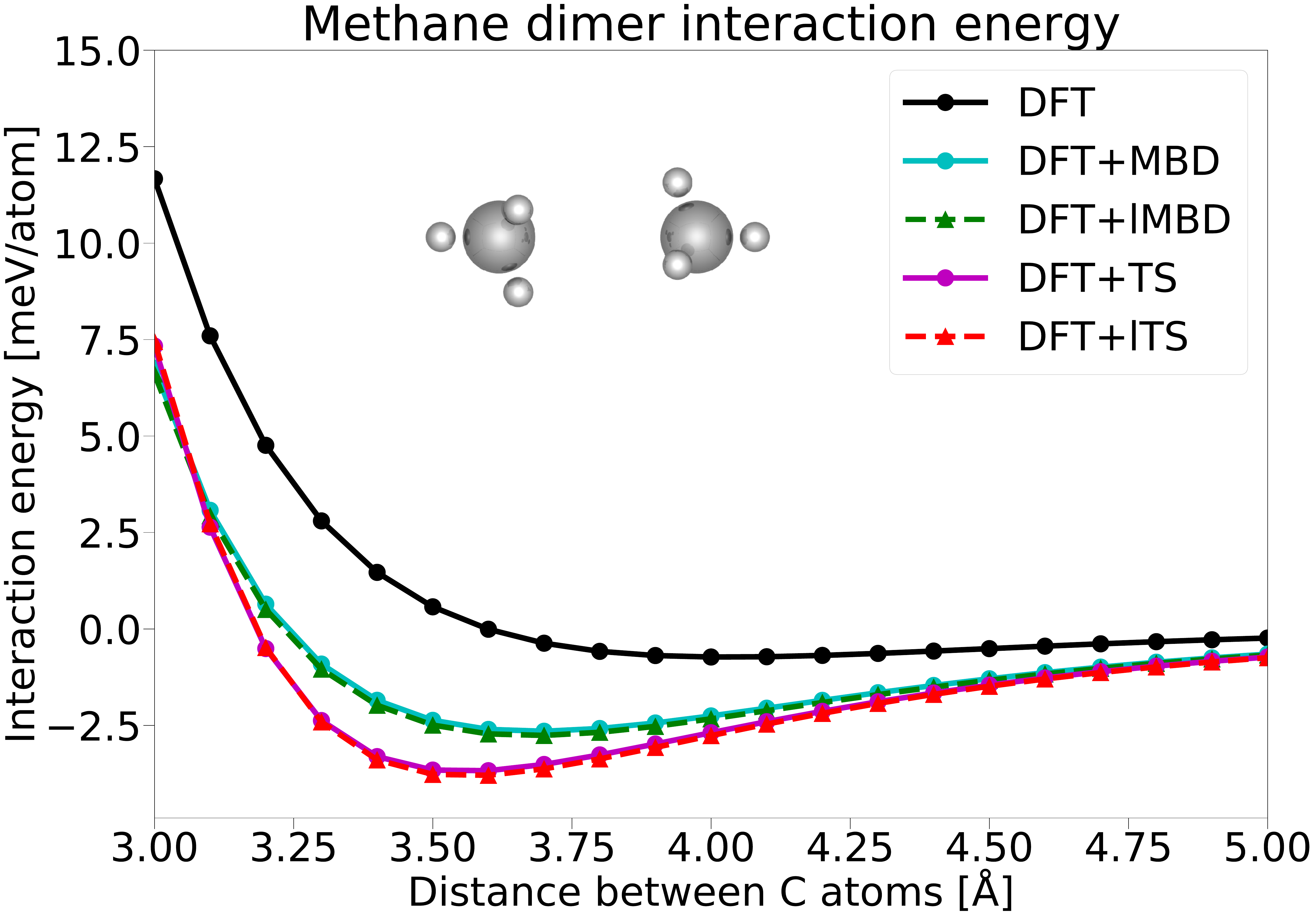}
&
\includegraphics[width=\columnwidth, trim=0.7cm 2.5cm 0.7cm 0cm]{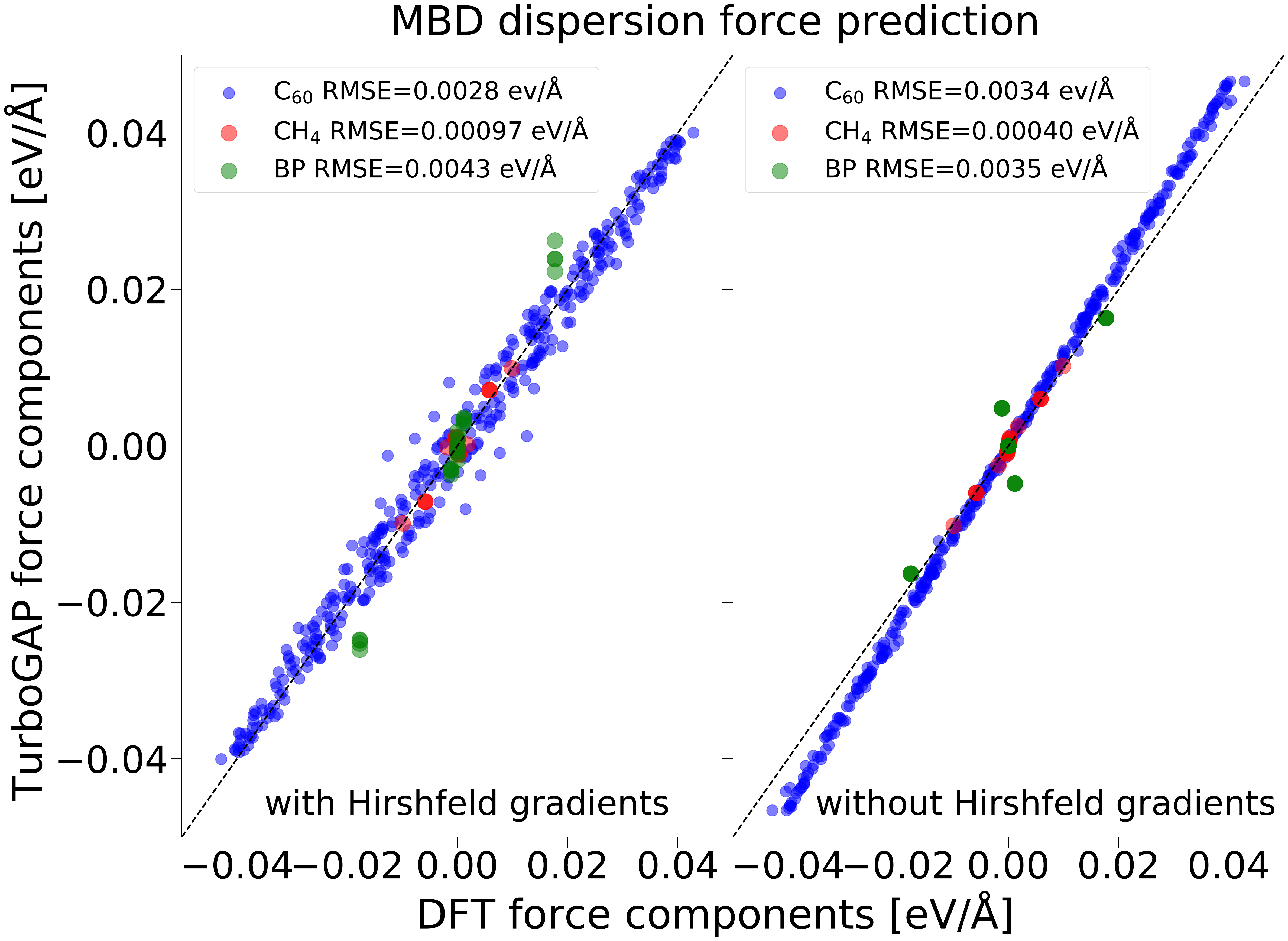}
\\
(c)
&
(d)
\\
\end{tabular}
\caption{Different comparisons between VASP's \gls{MBD} implementation and TurboGAP's lMBD implementation, as well as the respective \gls{TS} implementations for further reference. The PBE-DFT energies (labeled DFT) are computed with VASP with the convergence parameters given in the text. VASP's MBD- and TS-corrected values are simply labeled as DFT+MBD and DFT+TS, respectively, whereas the DFT energies corrected with TurboGAP's lMBD and lTS models are labeled DFT+lMBD and DFT+lTS, respectively. Note that we use DFT energies as baseline in all cases to avoid muddying the comparison by adding extraneous errors introduced by, e.g., using an ML force field to compute the dispersionless energies, as our focus is on comparing the different models of the dispersion corrections only. (a) C$_{60}$ interaction energy curve. The distance is between the centers of mass of the C$_{60}$ molecules. Note that, in the absence of dispersion corrections, PBE-DFT predicts no minimum, whereas the TS- and MBD-corrected curves both produce the minimum energy at around 10~Å distance. (b) Bulk black phosphorus exfoliation curve as a function of the lattice parameter in the interlayer direction. The unit cell is shown in the inset of the figure with the longest lattice vector corresponding to the interlayer direction. In this calculation the level of the dispersion correction matters as the optimal lattice parameter differs noticeably for the TS and MBD calculations. (c) Methane dimer interaction energy curve as a function of the distance between the carbon atoms. The orientation of the molecules is given in the inset. (d) Scatter plot of the lMBD forces calculated by TurboGAP and compared to the ones calculated by the DFT reference method. lMBD forces use the approximation given in Eq.~\eqref{cent-appr} while DFT forces neglect the gradients of the effective Hirshfeld volumes. For this reason, we have included a calculation with and without the Hirshfeld gradients (labeled in the figures). In the comparison with the gradients, the accuracy of lMBD seems fine for the methane (red) and C$_{60}$ dimers (blue), despite a few outliers. For bulk black phosphorus there seems to be a small systematic error and the comparison is noisy in general due to the lack of Hirshfeld gradients in the VASP calculation. The comparison where both VASP and TurboGAP calculations are done without the Hirshfeld gradients gets rid of the noise and reveals that there is a systematic error for C$_{60}$ because of the force approximation and a set of forces for bulk black phosphorus with small magnitude get an incorrect sign, which we also attribute to the force approximation.}
\label{c60_interaction}
\label{BP_exfol}
\label{methane_interaction}
\label{methane_dispersion_interaction}
\label{forces_scatter}
\end{figure*}

\begin{figure}
\begin{center}
\includegraphics[width=\columnwidth, trim=0.7cm 1cm 0.7cm 0cm]{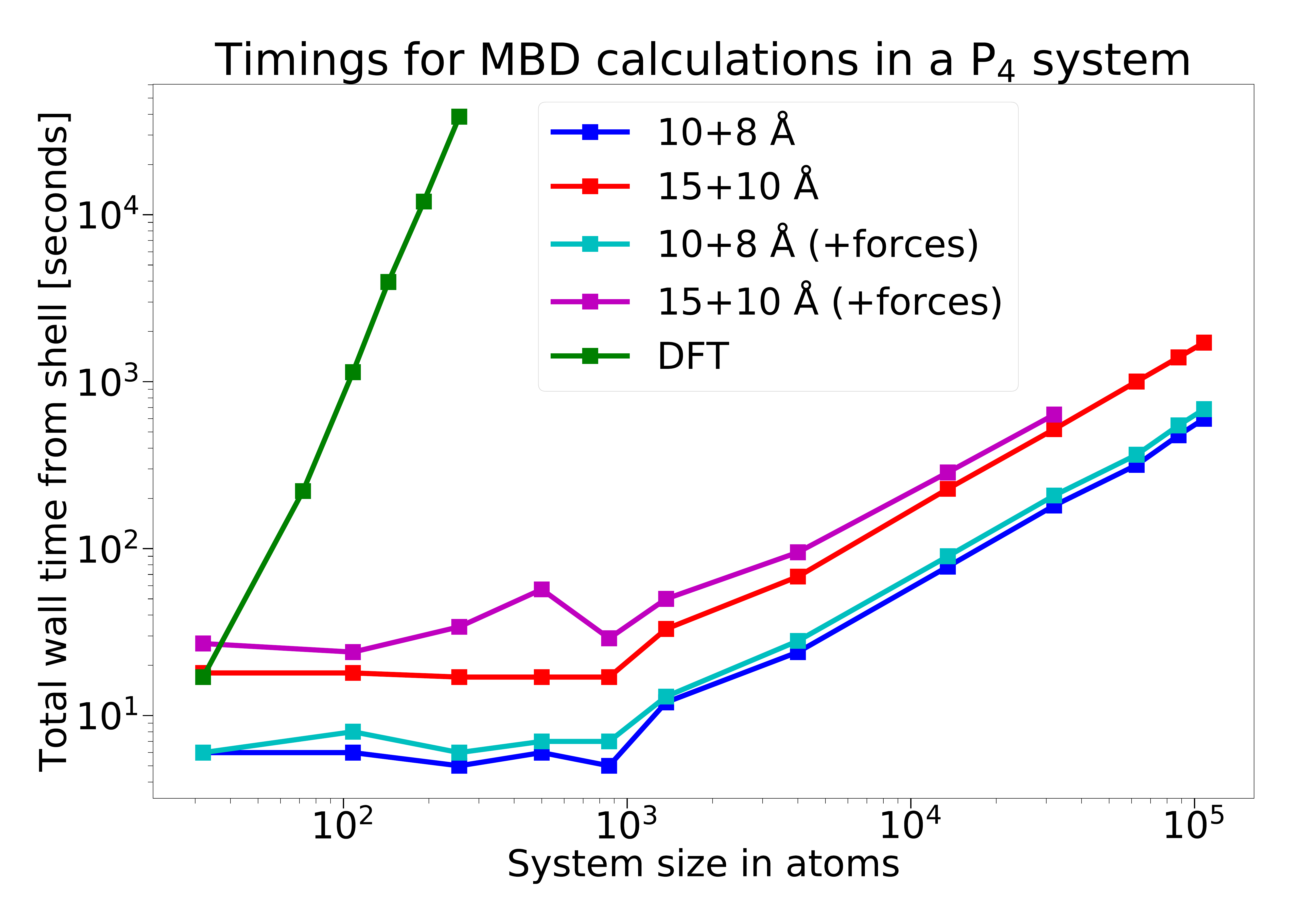}
\end{center}
\caption{The timing calculations were run on a system with randomly rotated P$_4$ molecules in a simple cubic lattice. All calculations used a total of 1024 cores and the timing of the base total energy calculation is subtracted such that only the contribution from MBD is shown. The legend shows the MBD primary + secondary cutoff radii with and without forces. The DFT calculation includes forces by default. DFT was run with VASP with optimal parallelization for our hardware (NCORE = 16 cores working on individual orbitals in parallel). The TurboGAP lMBD calculation was parallelized for one atom per CPU, which makes the timing curve flat before the number of atoms exceeds the number of CPUs (1024) in the calculation. After that, the lMBD scaling is linear. The DFT calculation seems to scale as $O(N^4)$, where $N$ is the number of atoms. The asymptotic scaling for \gls{MBD} should be $O(N^3)$ (and $O(N^2)$ for small systems)~\cite{hermann2023libmbd}. See text for a discussion.}
\label{timings}
\end{figure}

\section{Benchmarks}

We have calculated some benchmark results for our lMBD implementation with ML-based Hirshfeld volumes to show that it closely approximates the reference MBD method with DFT-based Hirshfeld volumes. We present results of interaction energies for various systems with different elements to show that the method is general and system-independent. Furthermore, we will show that the force approximation given in the previous section achieves sufficient accuracy and that the timing for running the calculations scales linearly with the number of atoms, demonstrating that the code is able to run calculations for much larger systems than typically accessible with DFT-based methods.

All lMBD energy and force calculations in this section were run with a maximum body order $n_{\text{max}}=6$ and the two-body cutoff radii $r_{\text{2b},1}$ and $r_{\text{2b},2}$ equal to the corresponding \gls{MBD} radii unless noted otherwise. The cutoff radii are given in each subsection. The DFT calculations were run with VASP with $k$-point sampling given by a parameter that defines the $k$-point density called KSPACING and it was set to 0.25~Å$^{-1}$. The energy cutoff for the plane-wave basis set, defined by a parameter called ENCUT, was set to 650~eV for C$_{60}$ and methane and 500~eV for black phosphorus.

\subsection{C$_{60}$ dimer}

First, we present the interaction energy curve for a C$_{60}$ dimer in Fig.~\ref{c60_interaction}~(a). The interaction energy for each point in the curve is calculated as:
\begin{equation}
    \Delta E(A,B) = E(A,B) - E(A) - E(B),
\end{equation}
where $E(A,B)$ is the total energy of the dimer including monomers $A$ and $B$ and $E(A)$ and $E(B)$ are the total energies of the same monomers individually, which are the same in this case. Each energy value is divided by the number of atoms in their respective system to get energies per atom. The TurboGAP calculations were run with a cutoff $r_{\text{SCS}} = 8.0$ Å for the polarizabilities and the MBD cutoff radii were set large enough to include all atoms.

In this calculation, we have included the bare DFT interaction energies to show how the system is inaccurately described without dispersion. These curves have their minimum at infinity which means that, in vacuum, C$_{60}$ molecules repel each other at every distance. This is not true in reality; the dimer has an optimal intermolecular distance where the repulsion matches the attraction. In the curves that include either TS or MBD dispersion correction, this optimal distance can be found as the minimum value of the interaction energy at around 10~Å distance between the C$_{60}$ centers of mass. For this particular system, the optimal distances obtained with MBD and TS are reproduced with lMBD and lTS, respectively. The main difference is that TS seems to overbind a bit compared to the more accurate MBD.

The small shift upwards for the DFT+lMBD curve compared to the DFT+MBD curve can be explained by the choice of cutoff for the maximum body order in Eq.~\eqref{E_MBD_k}: the VASP calculation uses the logarithm which is the limit of the series at infinite body order.

\subsection{Bulk black phosphorus}

Next, we present the black phosphorus exfoliation curve for TS and MBD dispersion corrections in Fig.~\ref{BP_exfol}~(b). Here we have only included the total energy per atom instead of the interaction energy of the system because we are looking at \textit{bulk} black phosphorus. For lMBD, the polarizability cutoff $r_{\text{SCS}}$ was set to 8~Å and the primary and secondary \gls{MBD} cutoff radii were set to $r_{\text{MBD,1}} = 20$~Å and $r_{\text{MBD,2}} = 14$~Å, respectively.

In this system, the choice of the dispersion method is important because the optimal lattice parameter in the interlayer direction shifts to a smaller value with \gls{MBD}, compared to TS. The minimum with TS is at around 10.8~Å and the minimum with MBD is at around 10.4 to 10.5~Å. The reported experimental value based on powder diffraction methods is 10.48~Å~\cite{brown1965refinement} so MBD improves the accuracy of the calculation. One can also see that the DFT+TS and DFT+lTS are almost on top of each other while DFT+lMBD diverges from DFT+MBD a bit (but retains the minimum) as the lattice parameter becomes smaller. The reasons for this are most likely the cutoffs for maximum body order and radius from central atom for DFT+lMBD: DFT+MBD effectively includes infinite body order and the cutoff for the system is defined by the $k$-point sampling (i.e., how many periodic replicas of the unit cell are included). As the system gets more dense, the eigenvalues of the matrix $\tilde{\mathbf{G}}_K(\omega)$ in Eq.~\eqref{E_MBD_k} and $\mathbf{A}_{\text{LR}}(\omega)\mathbf{T}_{\text{LR}}$ in Eq.~\eqref{E_MBD} usually get larger in magnitude, which in turn requires DFT+lMBD to get a larger maximum body order to match the DFT+MBD calculation. It is also difficult to compare the cutoff radius of lMBD to the $k$-point density of VASP's \gls{MBD} implementation as the approaches are fundamentally different. Nevertheless, lMBD seems to reproduce the \gls{MBD} result quite well and is definitely an improvement over TS.

\subsection{Methane dimer}

In the final energy benchmark we show the interaction energy of a methane dimer in Fig.~\ref{methane_interaction}~(c). The cutoff radius for the polarizabilities in the calculation was $r_{\text{SCS}} = 8$~Å and the primary and secondary \gls{MBD} radii were set large enough to include all atoms. In this example, the agreement between MBD and lMBD, on the one hand, and TS and lTS, on the other, is very good. We also note the importance of accurately capturing vdW interactions to correctly describe the physics of hydrocarbons and other weakly bonded molecular systems, as seemingly small energy differences might have a rather pronounced impact on, e.g., the density at those pressures most relevant to industrial applications. In addition to accurate vdW, these systems may also require the inclusion on quantum nuclear effects due to the presence of hydrogen atoms~\cite{veit_2019}.

\subsection{Force comparison}

We have calculated dispersion forces for lMBD, with and without Hirshfeld volume gradients, as given by Eq.~\eqref{cent-appr}. The forces were calculated for the C$_{60}$ dimer, bulk black phosphorus and the methane dimer discussed in the previous sections, using the same parameters as for the energy calculations, with the exception of black phosphorus where we reduced the secondary \gls{MBD} cutoff radius $r_{\text{MBD},2}$ from $14$~Å to $12$~Å. The MBD forces were calculated by running DFT+MBD and then bare DFT and subtracting the total forces to get only the contribution from dispersion. 

Based on the scatter plot, Fig.~\ref{forces_scatter}~(d), the approximation for the forces performs quite well. In fact, we note that the DFT-derived MBD forces have an intrinsic noisy contribution, due to neglecting the Hirshfeld volume gradients, that is similar in magnitude to the error incurred by the lMBD approximation. As mentioned earlier, we can seamlessly incorporate these gradients when the Hirshfeld volumes are obtained from an ML model~\cite{muhli2021machine}.

The most straightforward comparison between MBD and lMBD is in the absence of the Hirshfeld volume gradients in lMBD as it allows for a one-to-one analysis. In this case, we observe a small but noticeable systematic overestimation of the magnitude of the forces for the C$_{60}$ dimer. This is possibly due to the body-order truncation to finite body order in the MBD expansion. Another possible explanation is due to the intrinsic nature of our approximation: the atoms are close enough to be affected by the SCS calculation but we neglect the gradients of the SCS polarizabilities in the higher than two-body terms for the long-range dipole coupling tensor elements that are not directly coupled to the central atom. The disagreement does not seem severe and the speed-up for the calculation because of the approximation is significant: we only need to calculate a matrix-vector product $n_{\text{max}}-2$ times instead of calculating a matrix-matrix product $n_{\text{max}}-1$ times. Finally, we note that the obvious C$_{60}$ outliers on the left-hand-side panel of Fig.~\ref{forces_scatter}~(d) are actually due to the error in the regular DFT+MBD calculation incurred by neglecting the Hirshfeld volume gradients, as these are gone when we also remove these contributions from the lMBD calculation.

For CH$_4$, the agreement is quite good and does not merit further analysis.

For black phosphorus, which is the only bulk material in our benchmarks, the magnitude of the forces with Hirshfeld gradients would appear to be almost consistently overestimated by our approach; however, we recall that the reference DFT calculation is missing the Hirshfeld gradients and thus the comparison on the right-hand-side panel, where the disagreement occurs only for some of the data points, is more significant.
When the comparison is made where both calculations are missing the Hirshfeld volume gradients, a set of outliers arise. These outliers have their force sign reversed in the lMBD calculation vs the reference, although they are rather small in magnitude. After investigating the outliers by tuning the cutoff for the body order and the \gls{MBD} radii, we could not fix the sign. We decided to run a full-force calculation without the approximation to get as close to the VASP's conditions as possible and that fixed the sign for the outliers. Thus, we conclude that these outliers arise from the force approximation.

All in all, the force comparison with and without Hirshfeld volume gradients reveals that the errors incurred by neglecting these gradients in the reference VASP implementation, visualized as noise (vertical scatter of data) on the left-hand-side panel of Fig.~\ref{forces_scatter}~(d), are of the same order of magnitude as the errors incurred by the different approximations we have introduced in the derivation of the lMBD formalism. This highlights the accuracy of the lMBD approach in the context of the accuracy of the existing DFT-based \gls{MBD} implementations.

\subsection{Computational cost}

To compare the computational cost of our method to DFT-based MBD, we ran some timing calculations for a system consisting of P$_{4}$ tetrahedron molecules in a simple cubic lattice rotated randomly about their center of mass. The calculations were run to demonstrate the linear scaling of our method compared to the cubic scaling of the reference method in VASP. The results also demonstrate that, due to the atom-wise treatment of the dispersion interactions, we are not as limited by the system size as the DFT calculations. The timings can be seen in Fig.~\ref{timings}.

We used two different primary and secondary cutoff combinations to calculate the lMBD energies with TurboGAP. The first set of calculations used a primary cutoff $r_{\text{MBD,1}}$ of 10~Å and a secondary cutoff $r_{\text{MBD,2}}$ of 8~Å. The second set of calculations increased the primary cutoff to 15~Å and secondary cutoff to 10~Å. The maximum body order was six for all calculations. We also ran a separate calculation that includes forces to show how much time the forces add compared to just energies and to also show that the scaling is linear for the forces as well. The DFT calculations used the same parameters that we used for the black phosphorus calculation earlier, that is, ENCUT = 500 eV and KSPACING = 0.25 Å$^{-1}$. All calculations were run with and without MBD corrections and the timing of the calculation without MBD was subtracted from the timing of the MBD calculation to get only the contribution added by the correction.

All calculations were run in parallel on 1024 cores on the Mahti supercomputer in CSC (www.csc.fi). In the TurboGAP calculations one core was working on one atom. In the VASP calculations, the optimal setting for our hardware architecture of 16 cores working on an individual orbital was used (NCORE = 16). It can be seen that the timing for lMBD calculations remains almost flat until the number of atoms exceeds the number of cores due to idling and communication overhead and, after that, the scaling is linear. Forces add a small overhead on top of the energy calculations because some of the linear algebra has to be performed twice. Because of our approximation, the overhead remains small. The force calculation with $15+10$ Å cutoff radii crashed for the largest three structures because we did not allocate enough memory to store all the central polarizabilities and force arrays on each core. By default, Mahti provides 1.875 GB of memory per CPU core~\cite{CSC}. The memory per atom could be easily increased by making more cores work on an atom (MPI node undersubscription) but it would require running some of the calculations with a different hardware setup. We decided that this is unnecessary because the memory usage in our code can still be improved to reduce the memory requirements and the given timings already demonstrate the linear-scaling behavior. Although code optimization is not a central objective of this paper, we are currently working on improving its efficiency to enhance the performance of lMBD calculations.

Comparing the timings of our method to the timings of VASP, we noticed that the MBD implementation of VASP seems to scale as $\sim O(N^{4})$, $N$ being the number of atoms in the system. The method itself should scale as $O(N^3)$ and the matrix construction as $O(N^2)$~\cite{hermann2023libmbd} so, deducing from the limited points we have for the DFT timings, this worse scaling could be due to the contribution of hardware- and/or software-specific factors. This is also the reason why we could not finish a calculation for a system consisting of just 500 phosphorus atoms before the time limit of 36 hours allocated for the calculation caused it to crash. Note that the purpose of this analysis is not to benchmark the reference VASP implementation but to verify, in practice, the theoretical linear scaling of lMBD. The VASP calculation also includes forces.

\begin{figure*}
\begin{center}
\includegraphics[width=\textwidth, trim=0.7cm 1cm 0.7cm 0cm]{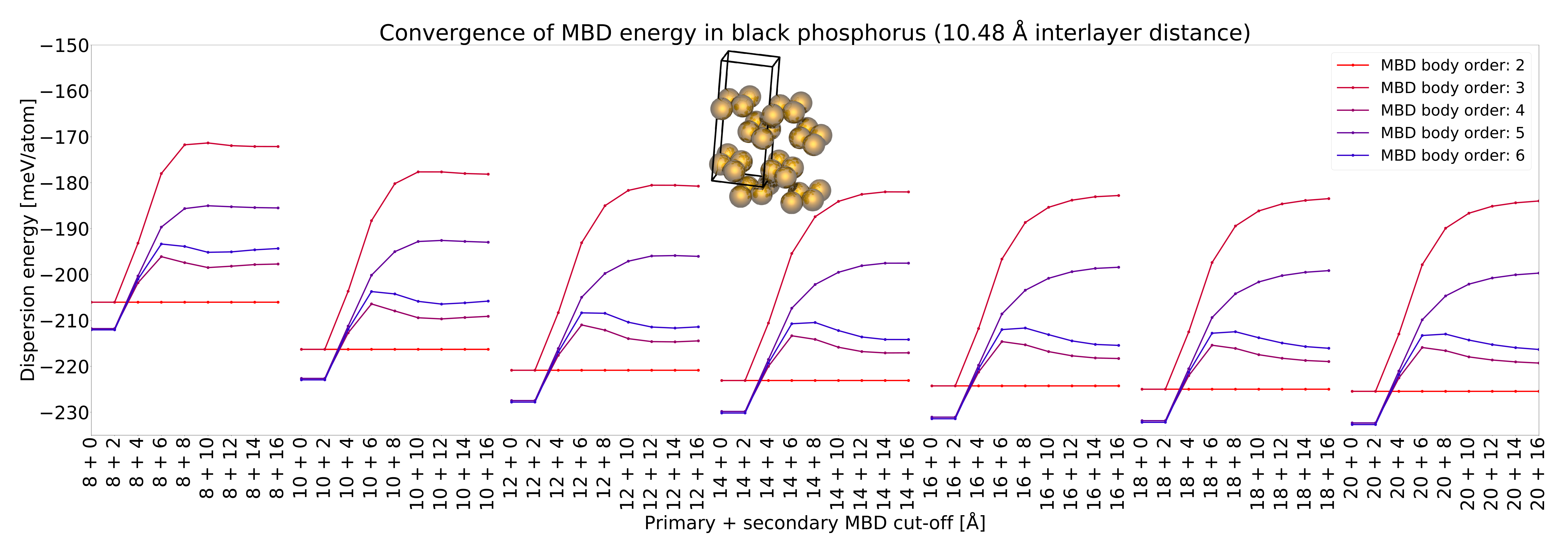}
\\
\includegraphics[width=\textwidth, trim=0.7cm 1cm 0.7cm 0cm]{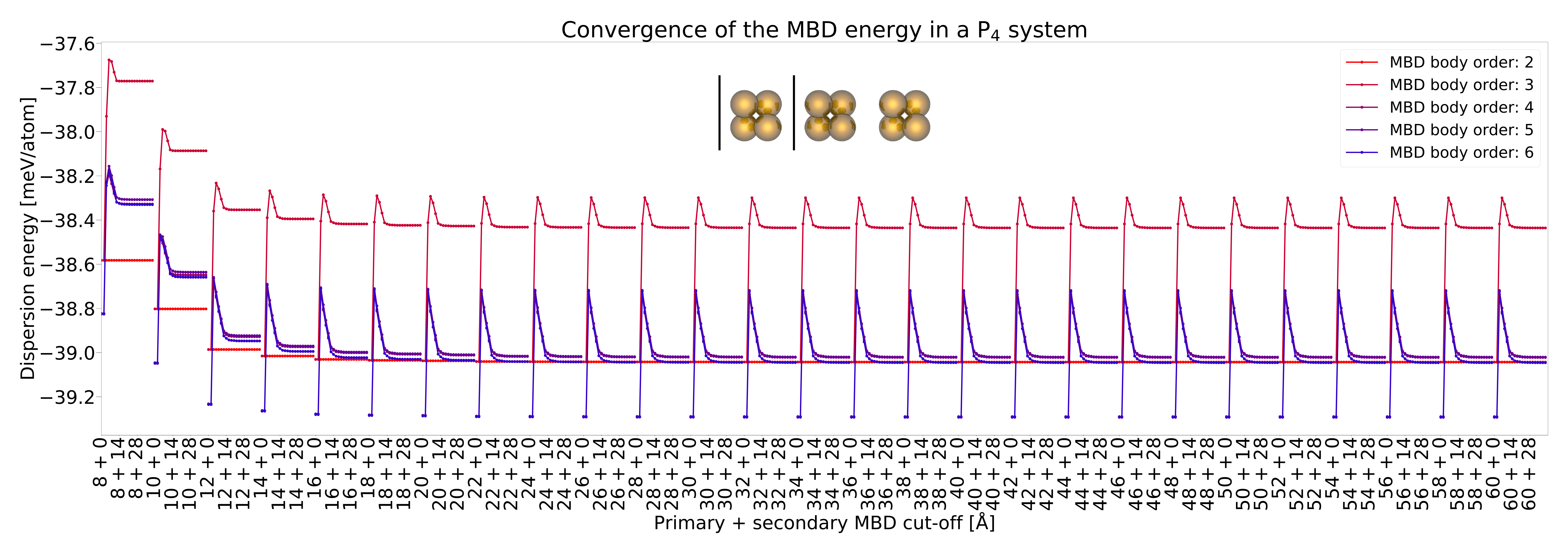}
\end{center}
\caption{Convergence results for (top) bulk black phosphorus and (bottom) P$_{4}$ molecules in a line. The vertical axis gives the dispersion energy and the horizontal axis gives the cutoff radii used in the calculation in the form primary + secondary cutoff in ångströms.
For black phosphorus (top), the unit cell of the material is shown in the inset.
For P$_4$ molecules (bottom), the vertical bars in the inset show the boundaries of the unit cell along the $x$ direction. The unit cell contains enough vacuum along the other Cartesian directions to remove any periodic replicas from the reach of the chosen MBD cutoff radii.
}
\label{BP_convergence}
\label{P4_convergence}
\end{figure*}

\begin{figure*}
\begin{center}
\includegraphics[width=\textwidth, trim=0.7cm 1cm 0.7cm 0cm]{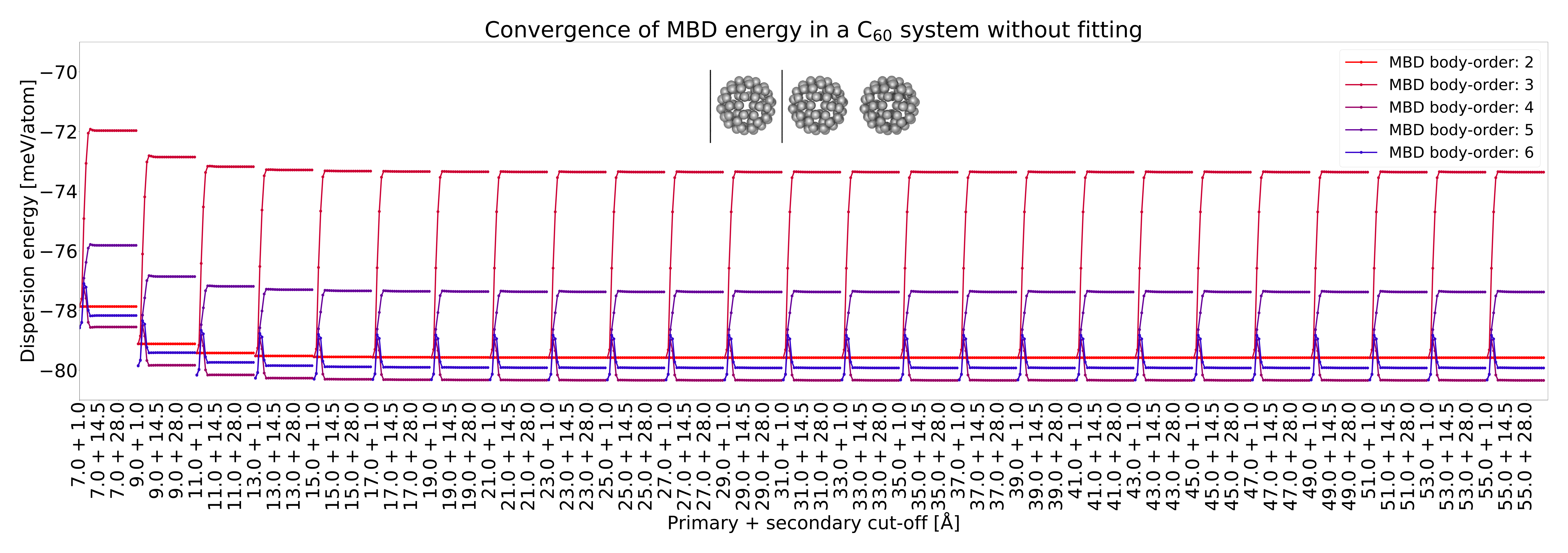}
\\
\includegraphics[width=\textwidth, trim=0.7cm 1cm 0.7cm 0cm]{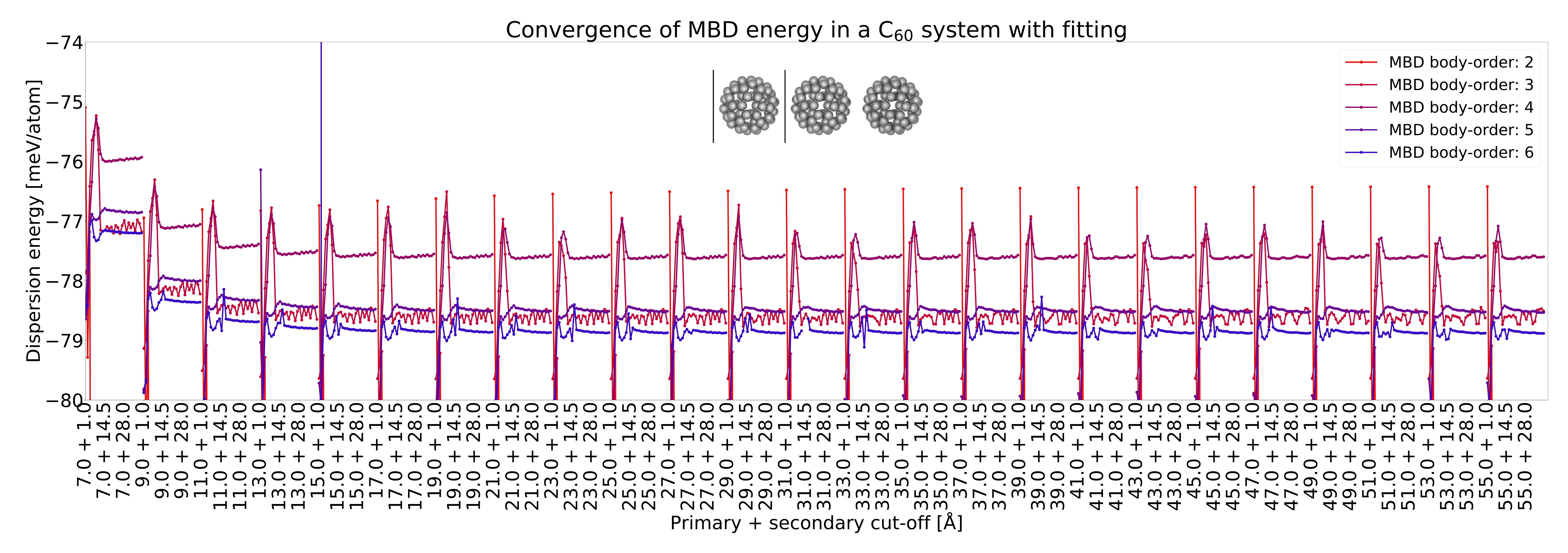}
\end{center}
\caption{Convergence results for C$_{60}$ molecules in a line.
The bottom panel shows the results using the power iteration approach to fit the logarithm, Eq.~\eqref{eigendecomp}, whereas the top panel uses the regular body-order expansion to approximate the logarithm, Eq.~\eqref{series-expansion}.
The vertical bars in the inset give the boundaries of the unit cell along the $x$ direction and the other Cartesian directions contain enough vacuum to bring the periodic replicas out of the reach of the MBD cutoff radii.
}
\label{c60_convergence_nofit}
\label{c60_convergence_fit}
\end{figure*}

\subsection{Convergence tests}

We ran tests for three different systems to show how our approach converges with respect to the cutoff radius and body-order truncation. The test systems were bulk black phosphorus, P$_{4}$ molecules in a line and C$_{60}$ molecules in a line. The first system was chosen because it is relevant to one of the benchmark interaction energy calculations and the latter two because molecules in a line allow one to study the effect of very large cutoff radii without running out of memory due to too many atoms within the cutoff sphere.

The convergence test for black phosphorus can be seen in Fig.~\ref{BP_convergence} (top). The vertical axis gives the calculated dispersion energy for the unit cell, divided by the number of atoms in the cell. The horizontal axis shows the lMBD cutoff radii in the form primary + secondary in ångströms, where primary refers to cutoff $r_{\text{MBD,1}}$ and secondary to cutoff $r_{\text{MBD,2}}$ in Fig.~\ref{mbd-radii}. The different colors are for different maximum body orders $n_{\text{max}}$ in Eq.~\eqref{E_MBD_k} and are denoted in the legend. We use the coefficients of the logarithm expansion (see Eq.~\eqref{series-expansion}) for $c_n$ in Eq.~\eqref{E_MBD_k} for now. Due to the higher atomic density in the system and thus higher memory requirements compared to the calculations with molecules in a line shown later, the range of cutoff radii is smaller in this test.

The two-body energy is a constant as a function of the secondary cutoff, which is to be expected as the neighbors of the central atom are not coupled to their own neighbors in the two-body approximation. Each primary cutoff sets the limit to which the energies converge as a function of the secondary cutoff for each body order. The convergence with respect to the body order, on the other hand, seems to oscillate between even and odd values around the true limit of infinite body order (equivalent to the logarithm given in Eq.~\eqref{E_MBD}). This infinite body order limit is not given in the convergence tests but can be seen in Fig.~\ref{convergence} for the C$_{60}$ dimer.

In Fig.~\ref{P4_convergence} (bottom) we have the convergence results of P$_{4}$ molecules in a line. The unit cell contains a single P$_{4}$ molecule, with extreme amount of vacuum along two of the Cartesian directions and a very short lattice vector along the perpendicular direction such that the molecules effectively form a line due to the periodic boundary conditions.
Contrary to previous results with the C$_{60}$ dimer and bulk black phosphorus, the dispersion energy does not seem to oscillate for even and odd body orders for this calculation. The energy first jumps up from the constant value of the two-body energy and then gradually decreases as a function of the body order and the secondary cutoff. The reason for this might be that the system has less degrees of freedom and is almost one-dimensional. 

Fig.~\ref{c60_convergence_nofit} (top) shows the convergence test for a similar system as the one in Fig.~\ref{P4_convergence} (bottom): here the P$_{4}$ molecule is replaced with a C$_{60}$ molecule and the smallest dimension of the unit cell is increased such that the molecule fits inside the cell. An interesting observation in this test is that the energy seems to oscillate a lot more as a function of the body order than it does as a function of the secondary cutoff radius. The direction of the oscillation seems to depend on the parity of the maximum body order again. The large oscillations might hint that the different body orders add significant contributions to the dispersion energy. The question of whether the absolute value of the dispersion energy is close to the limit (the logarithm, see Eq.~\eqref{E_MBD}) is often somewhat irrelevant as we are usually interested in the relative energies. It is important, however, that the series converges to this limit as we will discuss below.

Finally, Fig.~\ref{c60_convergence_fit} (bottom) shows the same convergence calculation for C$_{60}$ molecules in a line but this time we are not using the coefficients of the logarithm expansion in Eq.~\eqref{series-expansion} but we are solving for the minimum and maximum eigenvalues of matrix $\mathbf{A}_{\text{LR},K}\mathbf{T}_{\text{LR},K}$ through power iteration and then fitting new coefficients to produce the energies at the logarithm limit, using Eqs.~\eqref{eigendecomp} and~\eqref{E_k}.
The power iteration method is unstable if the secondary cutoff is zero, but otherwise the convergence seems to be faster as a function of the body order. One needs to remember that if the eigenvalue spectral radius (the maximum absolute eigenvalue) $\rho(\mathbf{A}_{\text{LR},K}\mathbf{T}_{\text{LR},K})$ exceeds one, then Eq.~\eqref{E_MBD} is no longer valid, because it is originally derived from the series expansion of Eq.~\eqref{series-expansion} that assumes this holds. If it does not hold, the series diverges, and it is pointless to fit another series to the logarithm that no longer works as the limit for the series. This method can, however, be used to reach the accuracy of the reference method faster as a function of the body order, if the absolute value of the dispersion energy is important in the calculations. The power iteration also requires more matrix-vector multiplications, which adds some computational cost.

\section{Summary and conclusions}

In the present work we have introduced a fast and accurate approach for many-body dispersion interactions in molecules and solids. This approach relies on an atom-wise calculation of the self-consistently screened polarizabilities and many-body dispersion energies that have previously been solved using a non-local ``whole system at once'' approach. Although the methodology presented here is an approximation by nature---since it is local---the resulting local polarizabilities and the sums of local dispersion energies give good agreement with their non-local equivalents for our representative test cases. Additionally, the local energies within this approximation are ``additive'' in the sense that the total energy can be written as a sum of atomic energies, defying the type-B non-additivity~\cite{dobson2014beyond} present in the reference method~\cite{buvcko2016many}, which, to our knowledge, has not been previously considered in this context. This local approach allows one to integrate the methodology into force-field calculations for accelerated \gls{MD} simulations whenever the Hirshfeld volumes used to parametrize the model can be provided for modest computational cost, e.g., by using an \gls{ML} model~\cite{bereau_2018,muhli2021machine}. Furthermore, because the non-local many-body problem is divided into smaller, local parts, the computational problem becomes linear scaling and fully parallelizable, allowing one to calculate dispersion energies for very large systems. The small energy scales of dispersion interactions might prove challenging for the optimization of sufficiently accurate force fields for the baseline energy, e.g., requiring purpose-specific potentials~\cite{veit_2019}. That said, our method does not depend on the framework used to derive the underlying potential energy of the system and can be used on top of standard \gls{DFT} just as well, especially considering that \gls{MBD} corrections can become the computational bottleneck in \gls{DFT} calculations for large systems.

The fundamental and inevitable drawback of this approach is that it is not well suited for small systems. This is because, in the case of small systems, the computational time and memory \textit{per atom} become higher than in the non-local approach where everything is solved for self-consistently in one calculation. However, the local approximation is necessary if one wants to be able to simulate systems with thousands or even millions of atoms.

While the computational complexity of popular implementations of the the original MBD algorithm is at least $O(N^3)$, as we have seen in the timing calculations, where $N$ is the number of atoms in the system, our implementation has the complexity of $O(N)$. The prefactor for each atom is of the order of the computational cost of matrix-vector multiplication, which is $O(N_{\text{neigh}}^2)$ for the naive version and even better for the modern algorithms and sparse linear algebra we are utilizing. Here $N_{\text{neigh}}$ is the average number of neighbors for each atom within their cutoff radii. The total number of matrix-vector multiplications directly depends on the maximum body order. However, it should be noted that this prefactor does not depend on the system size but only on the atomic density, and the implementation thus scales linearly with the system size, that is, the number of atoms $N$. The prefactor remains the same for the force calculation, because of our approximation for the polarizability gradients. The local matrix calculations can still become quite computationally expensive while the contributions from higher body-order terms might be negligible. Because of this, it is possible to run just the two-body version of this method which is equivalent to \gls{TS} corrections with \gls{SCS} polarizabilities~\cite{tkatchenko2012accurate,buvcko2013tkatchenko} (with different parameters) and much cheaper due to the manipulation that can be performed on the trace of a product of just two matrices. Thus, lMBD provides a seamless transition between \gls{TS}-\gls{SCS} and full \gls{MBD} within a physically transparent, systematically convergent framework for the inclusion of higher body-order and longer-range interactions.

A challenge for future work is the handling of cases where the series expansion in Eq.~\eqref{series-expansion} fails, i.e., the cases where the eigenvalues are too large in magnitude because of extremely small inter-atomic distances (or metallic systems). This has already been considered for the reference method~\cite{gould2016fractionally}, but our method requires a slightly different approach. Gould \textit{et al}.~\cite{gould2016fractionally} have also considered implementing a fractionally ionic approach for ionic systems (where \gls{MBD} produces larger errors)~\cite{nickerson2023comparison}. One option also worth considering would be to obtain the polarizabilities for \gls{MBD} using a more recent and more stable model, such as MBD-NL~\cite{hermann2023libmbd,PhysRevLett.124.146401}, provided that the method is suitable for machine learning the local parameters of the model. Furthermore, we also discussed the attractiveness of the \gls{XDM} method~\cite{becke2007exchange} in the Introduction due to its better computational scaling and good accuracy~\cite{nickerson2023comparison}. It would be interesting to explore whether the exchange holes used in \gls{XDM} are ``local enough'' for predicting them accurately within existing atomistic ML architectures.

\begin{acknowledgments}

The authors acknowledge financial support from the Research Council of Finland/Academy of Finland through grants numbers 321713 (H.M. and M.A.C.), 347252 (H.M. and M.A.C.) and 330488 (M.A.C.), as well as Horizon Europe's EuroHPC Joint Undertaking under grant agreement number 101118139 (Inno4scale, innovation study 202301-050, XCALE). T.A-N. has been supported in part under Academy of Finland’s grants to QTF Center of Excellence no. 31229 and European Union – NextGenerationEU instrument no. 353298.
Computational resources from CSC -- the Finnish IT Center for Science and Aalto University's Science-IT Project are gratefully acknowledged.

\end{acknowledgments}

\end{document}